# O VI IN THE LOCAL INTERSTELLAR MEDIUM

M.A. Barstow,[1] D.D. Boyce,[1] B.Y. Welsh,[2] R. Lallement,[3] J.K. Barstow[1,4], A.E. Forbes[1] and S. Preval[1]

[1] Department of Physics & Astronomy, University of Leicester, University Road, Leicester, LE1 7RH, UK
[2] Experimental Astrophysics Group, Space Sciences Laboratory, UC Berkeley, Berkeley, CA 94720 USA
[3] Service d'Aéronomie du CNRS, 91371, Verrières-le-Buisson, France
[4] Atmospheric, Oceanic and Planetary Physics, Department of Physics, Clarendon Laboratory, Parks Road, Oxford OX1 3PU, UK

ABSTRACT

We report the results of a search for O VI absorption in the spectra of 80 hot DA white dwarfs observed by the *FUSE* satellite. We have carried out a detailed analysis of the radial velocities of interstellar and (where present) stellar absorption lines for the entire sample of stars. In approximately 35% of cases (where photospheric material is detected), the velocity differences between the interstellar and photospheric components were beneath the resolution of the *FUSE* spectrographs. Therefore, in 65% of these stars the interstellar and photospheric contributions could be separated and the nature of the O VI component unambiguously determined. Furthermore, in other examples, where the spectra were of a high signal-to-noise, no photospheric material was found and any O VI detected was assumed to be interstellar. Building on the earlier work of Oegerle et al. (2005) and Savage & Lehner (2006), we have increased the number of detections of interstellar O VI and, for the first time, compared their locations with both the soft X-ray background emission and new detailed maps of the distribution of neutral gas within the local interstellar medium. We find no strong evidence to support a spatial correlation between O VI and SXRB emission. In all but a few cases, the interstellar O VI was located at or beyond the boundaries of the local cavity. Hence, any T ~ 300,000K gas responsible for the O VI absorption may reside at the interface between the cavity and surrounding medium or in that medium itself. Consequently, it appears that there is much less O VI-bearing gas than previously stated within the inner rarefied regions of the local interstellar cavity.

*Subject headings:* ISM: clouds – ISM: general – ISM: structure – ultraviolet: ISM – ultraviolet; stars – white dwarfs

## 1. INTRODUCTION

The existence of the diffuse soft X-ray background (Bowyer et al. 1968) implies that a substantial fraction of the Galactic disk, at least near the Sun, is filled with low density hot gas (McCammon et al. 1983). This discovery has important implications for our understanding of the structure and ionization of the interstellar medium (ISM) and in subsequent decades further studies have been made of the diffuse background, with a full sky survey being conducted by the *ROSAT* mission (Snowden et al. 1995, 1998). The total soft X-ray emission has been interpreted as the sum of three distinct components; extra-galactic, galactic halo and Local Bubble (LB) hot gas contributions. Details of these models and the constraints on the location of the soft X-ray emitting gas deduced from a number of soft X-ray shadows are discussed by Snowden et al (1998, 2000) and Kuntz & Snowden (2000), indicating that the temperature of the local hot gas is $T \approx 10^6$ K. If this gas is homogeneously distributed in the LB, the measured emission in any direction must be proportional to the path length through the hot gas to the surrounding dense, cool material. Therefore, it is possible to map the shape of the emitting region (Snowden et al. 1998). Among the main features are two large elongations towards opposite directions at high galactic latitudes. Their axis is tilted by about 20-30° with respect to the galactic polar axis, coinciding with neutral gas-free "chimneys" found nearly perpendicular to the Gould belt (Sfeir et al. 1999; Welsh et al. 1999).

Complementary to the study of the diffuse soft X-ray background are observations of absorption from the lithium-like O VI ion transition doublet in the far-ultraviolet at 1031.91 and 1037.62 Å, arising in the ISM (Rogerson et al. 1973; York 1974; Jenkins & Meloy 1974; Jenkins 1978). Ionization of O V to O VI requires an energy input of 114 eV. Since the local interstellar radiation field declines steeply at energies above this value (Vallerga, 1998), it is very difficult to account for the presence of O VI with a photoionization model. Therefore, the best explanation is that the O VI arises from collisional ionization in gas at a temperature in excess of $2 \times 10^5$ K. This proposition is supported by relative breadth of the O VI absorption line profiles, compared to those of lower ionization potential, which are consistent with thermal broadening at this temperature. In this paper we refer to the T~300,000K material which may be responsible for O VI absorption as "transition temperature" gas to differentiate it from the "hot" ~1,000,000K X-ray emitting gas.

The evidence for the presence of hot gas in the LB seems to be strong. However, there are a number of other observational results that contradict this picture. Any hot gas in the local ISM should give rise to a diffuse emission which should be detected not just in the soft X-ray waveband, but also in the extreme ultraviolet. Analysis of the *ROSAT* Wide Field Camera (WFC) EUV sky survey data reveals that the diffuse EUV background is negligible compared to other sources (West et al. 1994). Similarly, attempts to detect EUV line emission from the hot gas in deep exposures with the *EUVE* spectrographs yielded only upper limits to the flux (Vallerga & Slavin, 1998). Both the *ROSAT* WFC and *EUVE* telescopes were optimized for the detection and observation of point sources rather than diffuse emission. In contrast, the Cosmic Hot Interstellar Plasma Spectrometer (*CHIPS*) was specifically designed to detect the line emission from the putative hot interstellar gas component. Again, no background emission was detected, although very tight constraints were put on the possible plasma emission measure (Hurwitz, Sasseen & Sirk 2005) placing it an order of magnitude lower than implied by the soft X-ray fluxes. Either there is less hot gas than expected, and the soft X-ray background has another source, or the hot gas is significantly depleted in iron relative to other heavy elements, much more than reported elsewhere for hot gas.

One of the surprise results from the *ROSAT* mission was the detection of X-ray emission from the comet Hayakutake (Lisse et al. 1996), subsequently interpreted as arising from charge transfer of heavy solar wind ions with cometary neutrals (Cravens 1997). This solar wind charge exchange (SWCX) mechanism should also be experienced by neutral atoms of interstellar origin within the Solar System. De-excitation of charge-exchanged atoms should then generate diffuse soft

X-ray emission throughout the heliosphere (Cox 1998, Cravens 2000). Long term enhancements in the background observed by *ROSAT* appear to be correlated with Solar activity (Freyberg 1994, Cravens 2000), indicating that a fraction of the soft X-ray background (SXRB) arises from the heliosphere rather than the LB. Lallement (2004) finds that the fraction of SXRB emission attributable to our local cavity is substantially lower than previously thought.

Nevertheless, recent measurements of O VI absorption in the spectra of hot white dwarfs observed by the *FUSE* mission provide renewed evidence for the presence of significant quantities of hot gas in the LB (Oegerle et al. 2005, Savage & Lehner 2006). These authors report positive detections of O VI absorption along several local lines of sight but, as pointed out in these papers, there is the problem of possible contamination from photospheric material in the hottest white dwarfs making interpretation of the spectral measurements difficult. Indeed, *IUE* and *HST* high resolution UV observations of DA white dwarfs show that high species such as N V are found in the photospheres of white dwarfs down to surprisingly low effective temperatures. Originally, resonance lines of N V, C IV and Si IV detected in the *IUE* echelle spectrum of GD659 ($T_{eff}$ = 35,500 K), reported by Holberg et al (1995) were thought to be circumstellar in origin. However, a new measurement of the stellar radial velocity shows that these features actually arise in the photosphere (Holberg et al. 2000). The same resonance lines are also detected at the photospheric velocity of REJ1032+532 (Holberg et al. 1999, $T_{eff}$ = 46,330 K) and photospheric N V is seen in the *HST*/GHRS spectrum of REJ1614-085 (Holberg et al. 1997, $T_{eff}$ = 38,500 K), indicating that it is not possible to assume that high ionization species are not photospheric in the cooler (i.e. $T_{eff}$ < 50,000 K) hot DA stars.

An additional concern is the apparent absence of detections of other high ionization interstellar species in the UV spectra of white dwarfs. At gas temperatures sufficiently high to populate O VI, we might also expect to see N V and C IV. The ratio of the populations of N V and C IV to O VI will depend on the gas temperature, but for typical interstellar structures this lies somewhere in the range 0.1-0.4 (Indebetouw & Shull 2004), indicating that the N V and C IV resonance doublets should be present if O VI is detected. However, if the O VI lines are weak, the N V and C IV may fall below their detection threshold. From an observational point of view, there are no detections of interstellar N V or C IV along the lines-of-sight to any local white dwarf observed with *IUE* (Holberg Barstow & Sion 1998). Furthermore, careful radial velocity analyses of a smaller but higher signal-to-noise sample of DA white dwarfs studied with *HST*/STIS showed detections of both photospheric and circumstellar N V and C IV (Barstow et al. 2003a, Bannister et al. 2003), but no interstellar features.

Savage & Lehner (2006) take the reasonable position that if no stellar or circumstellar heavy element lines are detected in a spectrum then any detected O VI is interstellar in origin. Ruling out a photospheric component is straightforward, since there are lines of C III/IV, O IV, Si IV, P IV/V and S IV/VI that are readily detectable in the FUSE band. It is harder to be sure that there is no circumstellar component because there are no important resonance lines in the FUSE band other than the O VI. Fortunately, there are existing data from *IUE* and *HST* for a number of stars that can be used to confirm the absence of circumstellar material (e.g. WD1314+293, WD1254+223) and the work of Bannister et al. (2003) indicates that circumstellar components are only present when there is also photospheric material in a given star.

When stellar metals are present, following Oegerle at al. (2005) and Savage & Lehner (2006), the only way to unambiguously determine the nature of a detection of O VI in *FUSE* white dwarf spectra is to carry out a thorough analysis of the radial velocity of known interstellar and photospheric features for each individual star for comparison with the O VI velocity. Even then the difference between the interstellar and photospheric velocities must be greater or equal to the velocity resolution of the instrument. Few stars which have photospheric features have sufficient velocity separation between these and the ISM lines. Hence, even samples of several 10s of objects, as studied by Oegerle et al. (2005) and Savage & Lehner (2006) are whittled down to just a few good targets. Since the work reported in these papers was carried out, more observations have been added to the *FUSE* archive, increasing the number of stars available for study. Often multiple observations of a single target are available and can be co-added to improve the resultant signal-to-noise. Furthermore, the archival data are now available with an improved version of the *FUSE* pipeline (version 3.2, compared to version 2.05 and version 1.8.7 used by Savage & Lehner and Oegerle et al. respectively), which (hopefully) yields better quality data in terms of signal-to-noise and calibration.

It is therefore timely to re-examine the incidence of O VI in hot white dwarfs observed by *FUSE* and, where present investigate its nature. We report here on such a new study including a sample of 80 stars drawn from the *FUSE* archive. Of this group, we have detected O VI in 26 cases where there is either no confusing photospheric component or when the interstellar and photospheric velocities are well-separated. When we compare the distribution of stars in this study with neutral gas density maps of the local ISM it appears that the majority of interstellar O VI detections appear to be located at the periphery of or outside the local bubble. Hence, there is little evidence for the presence of million °K gas within the local bubble. In this paper we describe our analysis of the white dwarf sample in detail and consider the implications our results have for our understanding of the local interstellar medium and the origin of the soft X-ray background.

## 2. THE WHITE DWARF SAMPLE

We have surveyed all the hot DA white dwarfs ($T_{eff}$ >20,000 K) observed by *FUSE* to look for evidence of the presence of interstellar O VI absorption. The total sample comprises 95 stars and includes all public observations accessible in the *FUSE* data archive up to 2007 August 1. However, while there are many observations of cooler DAs and also a substantial number of observations of He-rich DO white dwarfs, we have excluded these targets from our study for various reasons. In the DA stars, at temperatures below ~20,000 K, the far-UV continuum flux is severely attenuated by the Lyman line series absorption in the spectral range occupied by the O VI doublet which yields a signal-to-noise below acceptable limits for all reasonable exposures. In some stars the continuum flux is contaminated by the presence of absorption by interstellar $H_2$, leading to confusion in identifying the lines we need to measure for this work or reducing the signal-to-noise of the data below useful levels. In principle, the DO white dwarfs, which all have temperatures above ~ 45,000 K, constitute excellent targets for this work. However, there are often many high ionization features present in these for which the atomic data for these is not very reliable making it hard to determine reliable radial velocities (see Barstow et al. 2007), in contrast to the DAs. In addition, these objects potentially contain significant abundances of Fe group elements with features distributed throughout the *FUSE* wavelength range which might significantly contaminate or mimic a weak detection of O VI. Our resulting target sample of 80 stars has considerable overlap with both those of Oegerle et al. (2005) and Savage & Lehner (2006), but is factors 3.6 and 1.7 larger, respectively. It is important to note that, because of the selection criteria we have outlined above; we have not included all the stars examined by these authors in this paper. On the other hand, they excluded from their analyses previously known stars where the O VI is thought to be contaminated or where the S/N is not adequate, which we do consider here.

All the white dwarfs we have considered in our detailed analysis are listed in Table 1, along with their physical parameters ($T_{eff}$, *log* g, mass and distance), where known, and their galactic coordinates. The targets include both isolated white dwarfs and also binary systems where the white dwarf has a main sequence companion (mostly those objects with an HD number). For several binaries, no published white dwarf temperatures and gravities exist, so we have estimated these values from the Lyman lines present in the *FUSE* spectra, following the methods outlined in Barstow et al (2003b). We have also updated some values where original estimates were made using just the Lyman α line in *IUE* SWP spectra. The list is ordered by the white dwarf RA/Dec number, where one has been assigned in the McCook and Sion (1999) catalogue, and any principal alternative name in general use is also given. Where there is no such name, the designation of the primary discovery survey is noted instead.

## 3. OBSERVATIONS AND DATA REDUCTION

### 3.1 *The FUSE Spectroscopic Data*

The FUSE mission was launched in 1999 and was successfully operated on-orbit until 2007. The instrument design and in-flight performance have been described in several scientific papers, including Moos et al. (2000) and Sahnow et al. (2000). The most recent version of the data processing pipeline, CALFUSE v3, is described by Dixon et al (2007). Briefly, the spectrographs are based on a Rowland circle design, comprising four separate optical paths (or channels). The channels must be co-aligned so that light from a single target properly illuminates all four channels, to maximize the throughput of the instrument. Spectra from the four channels are recorded on two microchannel plate detectors, with a SiC and LiF spectrum on each. Each detector is divided into two functionally independent segments (A and B), separated by a small gap. Consequently, there are eight detector segment/spectrometer channel combinations to be dealt with in reducing the data. Maintaining the co-alignment of individual channels has been difficult in-orbit, mainly due to thermal effects. A target may completely miss an aperture for the whole or part of an observation, while being well centered in the others. Hence, in any given observation, not all of the channels may be available in the data. To minimize this problem, most observations have been conducted using the largest aperture available (LWRS, 30 × 30 arcsec). This limits the spectral resolution to between 15000 and 20000 for early observations or 23000 later in the programme when the mirror focusing had been adjusted (Sahnow et al. 2000), compared to the 24000-30000 expected for the 1.25 × 20 arcsec HIRS aperture. Various spectra analysed here were obtained through HIRS, MDRS or LWRS apertures and in TIMETAG or HISTOGRAM mode. Table 2 summarises the datasets used for each of the white dwarf spectra studied in this paper.

### 3.2 *Data reduction*

All the data used in this work were obtained from the archive (see Table 2) after reprocessing using V3.2 of the CALFUSE pipeline. The data supplied by the archive consist of a set of files corresponding to the separate exposures for each of the 8 channel/detector segment combinations. Since the signal-to-noise of these can be relatively poor and the wavelength binning (~0.006 Å) over-samples the true resolution by a factor ~10, we have established a tried and tested to procedure to combine all the spectra (see Barstow et al. 2003b). Initially all the spectra were re-binned to a 0.02 Å pixel size for examination. Any strong interstellar absorption features present are used to verify that the wavelength scales for each exposure are well aligned. Individual exposures are then co-added to produce a single spectrum, weighting the individual spectra by their exposure time. This whole procedure was repeated for all eight detector/channel side combinations.

In principle, since we are mainly concerned with the analysis of the O VI lines in this paper, it is not necessary to study any segments other than the LiF 1A, SiC 1A, LiF 2B and SiC 2B, which encompass the O VI doublet. However, for this work we also need to make accurate measurements of the interstellar and photospheric velocities. Ideally, these would be simply obtained from photospheric and interstellar features immediately adjacent the O VI lines. Unfortunately, in many of the stars in our sample this is not practical, since there are no suitable lines. There may be no photospheric lines at all in the segment, or the interstellar lines are saturated and/or contaminated with geocoronal emission preventing accurate determination of the line velocity.

The perceived radial velocity of the photospheric lines will be the combination of both the intrinsic radial velocity of the star and its gravitational redshift, the latter depending on the stellar mass and radius. This could be measured at other wavelengths for our programme stars, but few reliable measurements are available for them in the literature. Both the *IUE* echelle and *HST*/STIS spectrographs had extremely good wavelength calibration and provide the best data but only ~25% of our sample have been observed by these instruments (see Barstow et al. 2003a, Bannister et al. 2003). Even in these stars, assigning a photospheric velocity to the *FUSE* data is a tricky issue because the *FUSE* wavelength scale zero point is not necessarily the same for each exposure and spectral segment due to flexure in the instrument and the general use of the largest LWRS aperture to compensate, which leads to a significant uncertainty in the positioning of a source within it. As a result the accuracy of the *FUSE* wavelength scale is typically limited to ~ 15 km s$^{-1}$. However, comparisons between the *IUE* echelle and *FUSE* for GD71 indicate that this uncertainty could be as large as ~20 km s$^{-1}$ (Dobbie et al. 2005). Therefore, we must be careful when directly comparing a velocity reference from *IUE* or *HST* with a *FUSE* measurement of an O VI velocity.

For the purposes of this analysis we do not actually need a very accurate absolute measurement of the radial velocities. What is most important is that we are able to obtain an accurate measurement of the *difference* between the interstellar and stellar components and an accurate determination of the O VI velocity on the same scale. Any undetermined offset to the wavelength zero point would be unimportant provided it is the same for all data segments. Therefore, to be able obtain good *relative* velocities it was necessary to combine the individual spectral segments to provide continuous coverage across the full *FUSE* spectrum. First, all the spectra were re-sampled onto a common wavelength scale of 0.02Å steps to reduce the level of oversampling. The re-sampled/re-binned spectra were then co-added, weighting individual data points by the statistical variance, averaged over a 20 Å interval, to take into account any differences in the effective area of each segment and any differences in exposure time that may have arisen from rejection of bad data segments (e.g. those with reduced exposure due to source drift). We have used the measured wavelengths of the strongest unsaturated interstellar lines to adjust the wavelength scale for each segment before co-addition, taking the LiF 1A as our reference point against which every other segment is adjusted.

Through visual inspection, it is apparent that the statistical noise tends to increase towards the edges of a wavelength range. In cases where the signal-to-noise is particularly poor (<3:1 per resolution element) in these regions, we have trimmed the spectra (by ~2-3 Å) to remove these data points prior to co-addition. An example of the resulting spectra, taking one of the higher signal-to-noise data sets, is shown in Fig. 1. The region of poor signal-to-noise seen in the 1080-1087Å region is due to the absence of LiF data at those wavelengths. Where we had multiple exposures for any white dwarf, we carried out a further stage of evaluation of the spectra, discarding those where the signal-to-noise was particularly poor or where the flux levels appeared to be anomalous, when compared to the remainder.

### 3.3 *Measuring photospheric and interstellar radial velocities*

The output of our data reduction procedure is a composite, wavelength calibrated spectrum covering the full range of the *FUSE* spectrographs for each white dwarf in the sample. The data for individual stars can then be visually searched for absorption features across the full wavelength range. In a separate exercise, we have cross-correlated the measured wavelengths of detected features with all the publicly available line lists (e.g. Kurucz and Bell, 1995; Morton 2003; Kentucky database – www.pa.uky.edu/~peter/atomic). A particular problem for line identification in the far UV is the large number of possible transitions for feasible species, often leading to several alternative nearby identifications (i.e. within the measurement and wavelength scale uncertainties) for a single feature. By requiring that all reliable identifications correspond to one of two possible radial velocities, either interstellar or photospheric, in an individual star and searching for similar absorption patterns in a number of stars, we have produced a master list for line identification purposes (Boyce 2009). This forms the basis for the determination of interstellar and stellar radial velocities in this paper.

For each star in the sample, we have searched for significant features across the full *FUSE* range and measured their wavelengths using a simple determination of the centroid of the feature by fitting a Gaussian profile. In general it is not necessary or appropriate to use more sophisticated line fitting procedures due to the under-sampling of the lines by the *FUSE* spectral resolution and the relatively low signal-to-noise of most features. The main purpose of this paper is to compare the observed velocity of any O VI features detected in these white dwarf spectra with the interstellar and stellar velocities to determine with which particular component, if any, it is associated. Ideally, to avoid any issues associated with differing wavelength calibrations for different spectroscopic segments, the comparison should be against lines that are close to O VI in wavelength and in the same spectral segment. The available interstellar reference lines are C II (1036.337), O I (1039.230 Å) and Ar I (1048.220Å), but there are few potential photospheric lines in the region apart from the O VI itself. Several P IV lines (1030.515, 1033.11, and 1035.516) fall in this wavelength range, but are present in only a few stars. Furthermore, any O I absorption may be contaminated by geocoronal emission and the C II can be saturated, compromising the wavelength measurement in either case. If we were to rely on velocity comparisons in the O VI region alone, we would have very few stars available for this study. Therefore, we have determined the mean interstellar and stellar velocities for each star from all the relevant lines available across the full *FUSE* spectral range. These results are summarized in Figure 2, Figure 3 and Table 3, together with the calculated velocity differences. Figure 2 compares the white dwarf photospheric heliocentric velocity with that measured for the ISM. There is no particular strong trend, but on average the photospheric velocities are offset by ~15-20 km s$^{-1}$, a value significantly lower than the 35.2±20.4 km s$^{-1}$ obtained by Savage and Lehner (2006). This is illustrated in Figure 3, showing the distribution of velocity differences ($v_{PHOT} - v_{ISM}$), which average 20.3±20.1 km s$^{-1}$. Within the measurement uncertainties, this is close to the 20 km s$^{-1}$ gravitational redshift of a typical 0.54 solar mass (R=0.017 R$_\odot$) hot white dwarf star.

To monitor consistency we have examined whether or not the velocities of individual lines vary systematically with wavelength and considered the scatter of the individual data points about the mean. Where we have reliable measurements of lines near O VI we have also compared these velocities with the computed means as a further consistency check. For all the stars in our study any discrepancies are well within the ~ 15 km s$^{-1}$ limit of the R ~ 20,000 *FUSE* resolving power.

Measurements of interstellar opacity made with a higher spectral resolution than is available with *FUSE* reveal that the interstellar lines of sight are often complex, with multiple components at different velocities (e.g. Sahu et al. 1999). Therefore, the interstellar lines observed in this work might well be blends of more than one component and the observed interstellar velocity the mean of these, weighted by the individual line strengths. The presence of unresolved blends may lead to increased scatter in the velocities we measure for individual ISM lines, as the weightings vary from line to line according to differing interstellar cloud conditions. However, in this work we are mostly interested in defining the separation between the ISM and photospheric velocities. The only real issue is that we need to take into account these systematic uncertainties in the analysis.

Table 3 lists the photospheric (where measureable) and interstellar velocities for each star in the sample, determined using the approach outlined above. We do not give statistical errors for these measurements because they are usually very small compared the systematic effects. The largest systematic issue is associated with the absolute velocity scale. However, any offsets apply equally to all our measurements and, as we are only interested in measuring the differences between the O VI, photospheric and ISM velocities, we can safely ignore this. In principle, the instrument resolving power gives an absolute limit (15 km s$^{-1}$, corresponding to the nominal LWRS resolution of 20,000) to the determination of velocities, but making the assumption that the observed lines should be symmetrical we can usually achieve a factor of ~ 3 better using a centroid determination algorithm rather than merely measuring the velocity of the minimum flux pixel in the line. Therefore, we assign a general uncertainty of 5 km s$^{-1}$ to all the present velocity measurements.

### 3.4 Detection and Measurement of O VI

We have inspected every *FUSE* spectrum in our sample to look for evidence of the presence of O VI, taking into account the possible uncertainties in its location, depending on whether or not the two possible lines at 1031.912 Å and 1037.613 Å might be located at the interstellar or photospheric velocity (see Tables 3 and 4). We also widened the search to ± 0.2 Å (60 km s$^{-1}$) either side of the range to ensure we did not miss any O VI that might be associated with another velocity component. O VI 1031.912 Å is detected in 41 objects in our sample, ≈ 50 % of the objects studied, and the weaker 1037.613 Å line is seen in 16 objects. The *FUSE* spectra in the vicinity of the O VI line detections are shown in Figure 4 for the 1031.912 Å line.

The 1037.613 Å line is detected only when 1031.912 Å is present, which is expected from the relative oscillator strengths. The issue of the significance of a detection of either O VI line is an important one, which obviously depends on the intrinsic line strength and the signal-to-noise of the data. In many white dwarfs the O VI lines are strong, and, therefore, obvious in the data, with their location well determined. However, weaker lines, particularly those close to the noise limit, are not so readily identified and separating real O VI detections from "noise" features becomes problematic. We have measured the velocities of all detected O VI lines as described in section 3.3 above. We also determined their equivalent widths (EW) in the standard way, as the area of the line falling below the nominal continuum flux level. The associated uncertainty in the equivalent width is estimated from the size of adjacent noise features in the spectrum, which allows us to assign a measure of significance for each "detection". Where there is no clear detection, we have determined a 3σ upper limit to the line strength from the typical size of noise in the vicinity of the O VI location.

Table 3 includes all O VI 1031.912 Å measurements, giving the measured wavelength, calculated radial velocity of the line, the equivalent width, its uncertainty and its significance (the equivalent width divided by the uncertainty). However, where only an upper limit to the O VI line strength is obtained, we only give that equivalent width as the other parameters are not relevant. Of the "detected" lines, the measured significance ranges from a maximum ~18 σ down to our 3 σ threshold. Any detection below 3 σ is not considered to be significant and is excluded from further analysis and discussion, apart from as an upper limit to the amount of O VI present. Similar data for the O VI 1037.613 Å detections are give in Table 4. All the upper limits for detection of O VI 1031.912 Å are listed in Table 3 but we provide no upper limits for the non-detections of the 1037.613 Å line because these far outnumber the reliable detections and, since they are calculated from the general noise level in the spectrum, will be the same as the values listed for the 1031.912 Å line.

A number of stars in the sample show evidence for multiple stellar components. Where these are clearly detected we note the velocity of the additional component in Table 3. There is no clear evidence of any secondary O VI component associated with any of these weaker features. However, in WD0445-282, the interstellar O VI absorption is broad and spans the weaker feature. There is also a hint in figure 4 that there might be a secondary O VI feature in WD1725+586, but it does not formally exceed our applied 3σ threshold. We discuss the particular properties of some individual stars and lines of sight in the Appendix.

### 4. DISCUSSION

### 4.1 Determination of the Origin of the O VI

As described earlier in this paper, there is strong evidence from other observations that relatively high ionization stages are seen in white dwarf photospheres at temperatures below those limits expected from stellar atmosphere calculations, due to stratification of material in the stellar envelopes. Therefore, when any metals are present in the atmosphere of a white dwarf we are unable to rely on ionization stage and stellar temperature as a means of discriminating between a photospheric and interstellar origin for the O VI. The only completely reliable method we can then use is to compare the O VI radial velocity with the values obtained for the photosphere and ISM in each star.

How well we can discriminate between a photospheric and interstellar origin depends on the separation of the two velocity components and the uncertainty in the velocity measurements. In section 3.3, we estimated that the systematic and limiting uncertainty in the velocity measurements was ~ 5 km s$^{-1}$. Hence, the photospheric and interstellar components must be at least separated by this amount for us to be able to discriminate. However, this is a somewhat naïve and likely underestimate of the magnitude of velocity difference that can be excluded with high confidence. Firstly, we are looking at velocity differences and, therefore, the combined uncertainty of two velocity measurements needs to be taken into account. Furthermore, the velocity measurement error adopted corresponds to a ~ 1 σ value. Therefore, there is a 68% probability that the true velocity lies within the quoted range of uncertainty. If we are to claim that a particular O VI line resides at either photospheric or interstellar velocity, we need to be able to rule out an association with the other component at high confidence. Adopting a 2 σ (10 km s$^{-1}$) uncertainty yields a 95% confidence limit. For the purposes of deciding the nature of the O VI components we look for a 14 km s$^{-1}$ (√2 x 10 km s$^{-1}$) minimum separation to rule out an association.

*Detection of photospheric O VI.* – As discussed in the introduction O VI can be present in white dwarf photospheres and is not necessarily restricted to high temperature objects. The detections are only designated photospheric if the detected line resides close to the measured stellar velocity and is well separated from the location of the expected interstellar component. Good examples are WD0131-164 or WD2211-495.

Slightly problematic are objects such as WD1029+537. In this example, the core of the O VI line is 1-2 wavelength bins redward of the photospheric velocity. However, this remains within the overall measurement uncertainties and so this is also classified as a likely photospheric detection. In any case, there is sufficient ambiguity to rule it out as interstellar in the context of this analysis. In many stars, we have a clear detection of O VI but the velocity difference between the ISM and photosphere is insufficiently large to make a clear statement about the nature of the detection. For example, in WD0501+524, the two components are virtually coincident, while in WD1611-084 they are separated by ~15 km s$^{-1}$, but still too small, when compared to the **width** of the line and the accuracy of the velocity measurements, to assign an origin. In the hottest white dwarfs it is most likely that the O VI is photospheric in origin. Below ~45,000-50,000K this is not necessarily the case, but since we know of several examples (WD0050-332 is one) where highly ionized metals are present at lower temperatures we cannot be definitive about the origin of the O VI.

*Detection of interstellar O VI* – Any detection recorded at a velocity other than the photospheric value we attribute to interstellar material. In 43 out of 80 white dwarfs a thorough search of the *FUSE* spectrum reveals no detectable photospheric material. If the O VI were photospheric, we would expect to see evidence of some other species. Therefore, we conclude that any O VI present is most likely to be interstellar in these stars. For most detections, the velocity of O VI was coincident, within measurement uncertainties, with that of the cooler ISM, as determined from neutral or singly ionized species. However, in several cases the O VI velocity was redshifted with respect to the ISM, as illustrated by WD1057+719 and WD1040+492. In one star, WD0455-282, the O VI is blueshifted.

*Detection of both interstellar and photospheric O VI* – Interestingly, there is also a clearly resolved photospheric component in the spectrum of this WD0455-282. There may also be a weak component at the ISM velocity, but this cannot be resolved from the broad blue-shifted feature. There are three other objects where both interstellar and photospheric components appear to be present. However, the separation of the ISM and photospheric velocity components is not large enough to completely resolve these. For example, in WD2321-549 exhibits a broad feature which appears to have two minima aligned with each velocity component. If there were no interstellar component or if it was much weaker than the photospheric contribution, the features would be strongly skewed towards the photospheric velocity. Hence, this is a clear indication that this is a blend of two lines. Three other stars, WD0027-636, WD2331-475 and WD2350-706, show similarly broad lines, which are good evidence for two blended components, but with less distinct separation. For these stars we record the full strength of the line from both contributions in Table 3 but take into account the photospheric contamination, which is approximately 50% of the total in these cases, by fitting an additional Gaussian profile in estimating the interstellar column and volume densities in Table 5. As a result, there will be some additional systematic uncertainty in the measurements. We have not included this in the error estimates but indicate the increased uncertainty in Tables 5 and 6.

Taking all the above into account we label the most likely origin of the O VI in Table 3 as: definitely or possibly photospheric (P), interstellar at $v_{ISM}$ (I), interstellar red-shifted relative to $v_{ISM}$ (IR), interstellar blue-shifted relative to $v_{ISM}$ (IB), not detected (ND). The interpretation of Savage & Lehner (2006) is also shown for comparison with our results. It is important to note that when we used the designation "P", we are indicating that the star contains other photospheric components which make it likely that there is a significant photospheric contribution to the O VI feature. However, this does not definitively exclude the possibility of an interstellar component being present, hidden by or blended with a photospheric contribution when the velocity separation is below the instrument resolution. In four stars, (WD0027-636, WD2321-549, WD2331-475 and WD2350-706), we see broad O VI components which we interpret as blended photospheric and interstellar lines and designate "I+P", even though not fully resolved. For these stars, the photospheric and interstellar components are severely blended and for our estimate of the interstellar contribution, we make the assumption that the photospheric and interstellar components are roughly equal in strength. The result is of course very uncertain and that is why we do not estimate any errors and use the symbol "~" to indicate the high level of uncertainty in Tables 5 and 6. In other examples, where the photospheric feature is dominant the presence of any ISM component is much more ambiguous and we are not able to carry out the same analysis.

4.2 *Comparison with other studies*

As discussed extensively already in this paper, there have been two previous studies of interstellar O VI using white dwarfs as background sources, by Oegerle et al. (2005) and Savage & Lehner (2006). Essentially, the Oegerle et al. sample is a subset of that of Savage & Lehner. Therefore, in this section we will just compare our results with the more recent paper. There is considerable agreement between the analysis we have conducted here and that of Savage & Lehner. We have 26 detections of interstellar material, compared to their 21 for the stars that are in common with this study. Since we apply a higher ($3\sigma$ cf. $2\sigma$) detection threshold a few detections reported by Savage & Lehner (WD0603-483, WD1019-141, WD1603+432 and WD1636+351) are not included in our list, although we do obtain similar equivalent width and significance in each case. There are two stars in the Savage and Lehner sample (WD0027-636 and WD2321-549) where they designated the O VI as photospheric but which appear to have a blended ISM component (also noted by them) and where we have estimated the strength of the ISM component.

Despite the fact that the sample of DA white dwarfs studied in the work is almost a factor two larger than that of Savage & Lehner, we have detected interstellar O VI in only 8 additional lines of sight an ~25% increase when counting just the $3\sigma$ and above detections, indicating a rather lower "success rate" in detecting interstellar O VI. Three of these (WD0027-636, WD2331-475 and WD2350-706) are also apparent blends of ISM and photospheric contributions. Hence, this may be partly due to Savage & Lehner's a priori exclusion from their analyses previously known stars where the O VI is thought to be contaminated or where the S/N is not adequate, which we did consider here.

### 4.3 *O VI column and volume densities*

We have reliable temperature and gravity estimates for all the white dwarfs in our sample, obtained from either Balmer or Lyman line analyses, as listed in Table 1. Coupling this information with visual magnitude and/or flux data allows us to estimate a distance to each object. This process depends on the assumption that the standard mass-radius relation for a typical DA white dwarf is correct (See Barstow et al. 2005, for further discussion on this topic) but with that, and the additional proviso that there are no hidden companions, yields distances accurate to a few percent (Holberg, Bergeron and Gianninas 2008) and which we have listed in Table 1. Consequently, we can examine both the distribution of column and volume densities for the O VI detections and upper limits. The upper limits obtained are listed in Table 3 and span a range of equivalent widths from 4.0 mÅ for the very highest signal-to-noise spectra to 113 mÅ, where the data have poor signal-to-noise. Since all the detected O VI lines are relatively weak, it is reasonable to assume that they are not saturated. Therefore, column densities have been calculated using the standard linear curve-of-growth method for unsaturated lines:

$$N = W_\lambda / (8.85 \times 10^{-21} f \lambda^2)$$

Where: N is the column density ($cm^{-2}$); $W_\lambda$ is the equivalent width of the observed line/upper limit (mÅ); f is the oscillator strength of the line (0.133 for O VI at $\lambda = 1031.93$Å). The average volume density along the line of sight (n, $cm^{-3}$) is then calculated simply by dividing by the distance to the star. The results of these calculations are summarised in Table 5 and in Figures 5 and 6, which show column and volume density as functions of stellar distance, respectively. The black diamonds and error bars are the measured values while the grey diamonds are the upper limits. For a number of stars with the highest signal-to-noise the implied densities determined from the upper limits have lower values than the actual detections. Hence, these limits provide tight constraints on the amount of hot gas present along those particular lines of sight.

One possible model for the Local Bubble is that it is full of hot X-ray emitting gas. If this material were uniformly distributed, we would expect to see O VI detections exclusively from transition temperature gas at the interface with cooler material, near the bubble boundary. Closer examination of Figure 5 shows that several of the well-constrained upper limits are obtained for stars that are closer to the Sun than the main group of detections. However, at least half the upper limits are at similar distances to stars with interstellar detections. Indeed, one non-detection (WD0229-481) is the most distant object in this group. This is a strong indication that the distribution of material around the Sun is far from uniform and that interfaces are not seen in all directions.

We already know from earlier studies of the LISM, culminating in the work of Lallement et al. (2003), that boundary of the Local Bubble is not equidistant from the Sun in all directions. Nor is it a simple shape. So perhaps it is not surprising that there is significant variance in the measured column densities and upper limits. For sources within the Local Bubble, we might expect the conversion to volume density to produce a more coherent picture. In fact, as can be seen in Figure 6, this is not so, with very low column densities being recorded over some of the larger distances. Again, the picture is one of extreme patchiness, with volume densities varying by an order of magnitude between stars at similar distances.

### 4.4 *The spatial distribution of the local O VI absorption*

Previous authors have already described the distribution of O VI absorption within 200pc of the Sun as being "patchy" (Oegerle et al 2005, Savage & Lehner 2006). If O VI forms at the evaporative interfaces between cool clouds and the surrounding hot million degree (soft X-ray emitting) Local Bubble gas, then the presence of tangled or tangential magnetic fields could quench thermal conduction over a large portion of a cloud's surface and thus O VI would only be formed in "patches" (Cox & Helenius 2003). However, Welsh & Lallement (2008) have suggested that many sight-lines along which local interstellar O VI absorption (d < 100pc) and O VI emission have been detected seem to preferentially occur at high galactic latitudes, in regions that also possess high levels of SXRB emission. To test this possible spatial correlation claim further with our new data, in Figure 7 we show the galactic distribution of the white dwarf sight-lines *within 120p*c together with B star sight lines (also *within* 120pc) listed by Welsh & Lallement (2008) and Bowen et al (2008) towards which O VI absorption has been detected with confidence, together with those sightlines (within the same distances) where we have measured only an upper limit to the possible O VI absorption strength. In Figure 7, these OVI data are superposed on a plot of the galactic distribution of the SXRB as mapped with the ROSAT satellite by Snowden et al (1998). For the purposes of this discussion we define areas of `enhanced SXRB emission' as those with a flux of > 650 R12 cts.

An initial inspection of the distribution of OVI detections presented in Figure 7 seems to confirm the previous claim of a "patchy" distribution. However, a deeper inspection of the data reveals the following. In total, Figure 7 contains 49 sight-lines, 15 of which have confirmed detections of interstellar OVI absorption. Of the 15 detections only two sight-lines are not coincident with directions where the soft X-ray background signal is enhanced over the lowest levels of SXRB emission. This seems to imply that whenever OVI is detected it tends to be in regions of enhanced SXRB emission. However, there are also 26 sight-lines located in regions of enhanced SXRB emission where OVI was not detected. Therefore, the presence of enhanced SXRB along a sight-line alone is insufficient to explain why OVI is preferentially detected.

In addition, of the 6 O VI detections located at mid-plane latitudes (-30 < b < +30 degrees), all are placed beyond 80pc (i.e. located presumably beyond the neutral boundary to the Local Bubble). This suggests that at mid-plane latitudes there is a deficiency of OVI bearing gas within the confines of the Local Cavity. The remaining 9 O VI detections (regardless of distance) are all located at high galactic latitudes (|b| < -30 and b| > 30 degrees), all located in regions of enhanced SXRB flux. However, there are also 15 non-detections at high latitudes that are located in regions of enhanced SXRB flux. Therefore, the location of a sight-line at high latitudes is in itself insufficient to explain why OVI is sometimes detected and sometimes not.

We note that recent calculations on the spatial distribution of the SWCX by Koutroumpa et al (2009) have suggested that the Local Bubble may contain far cooler and more tenuous gas at low latitudes, with the SXRB at high latitudes arising from an external hot and highly ionized gas located in the overlying galactic halo (Welsh and Shelton 2009). Although Figure 7 only very tentatively supports a scenario in which interstellar O VI could be formed in a transition region between hot (halo) gas and lower temperature (Local Bubble) gas at high latitudes, the presence of the (albeit few) O VI detections at low latitudes requires an alternative explanation. This is now provided by our interpretation of Figure 8.

Figure 8 indicates the location of a selection of the stars listed in Table 3 and shows a number of slices through the local ISM based on the 3-D spatial distribution of neutral gas presented by Lallement et al (2003), but updated by Welsh et al (2010) with more recent information. On each of the projections shown, we have plotted the positions of detections of interstellar O VI along with the most sensitive non-detections. Each map shows the neutral gas density in a pseudo-grey-scale, white representing low density and black high density, perpendicular to the galactic plane. All distances are in parsecs and the longitude of each slice marked along with the north and south galactic poles. All stars are marked with the log of their O VI column density and a number, for clarity, with the key to this labeling given in Table 6. Stars marked with filled circles correspond to ISM detections while the open circles represent non-detections. One star (21), a non-detection of O VI, lies at a distance above 200pc and is outside the boundary of the maps. Its direction is indicated by the arrow and distance noted on the plot. It is important to note that the density distributions of Lallement et al. (2003) have been obtained by inversion of a limited number of sight-lines. As a consequence their spatial resolution is coarse (of the order of 25 parsecs) and there may be clouds with a smaller extent that have not been detected and included in these maps. However, within these limitations, they give an idea of the ISM distribution along the lines of sight to the white dwarfs in this study.

Figure 8 shows that for the 5 white dwarf sight-lines located well within the inner regions of the Local cavity (i.e. white regions in Figure 8), we find only 2 O VI detections (towards targets 31 and 32), the nearest detection being towards WD2004-605 (target 32) at a distance of 58pc. Again, this supports the notion that there is a deficiency of O VI-bearing gas within the Local Cavity. Figure 8 also shows that the majority of white dwarfs yielding interstellar detections of O VI are mostly located close to or beyond the neutral boundary to the Local Cavity in regions of higher gas density (darker shaded regions in Figure 8). Some O VI detections lie at high galactic latitude, including most of the red-shifted components (as noted by Savage and Lehner 2006).

We note that in order produce O VI an interface between hot (million K) and cooler gas is required. Since the Local Bubble is known to possess numerous partially ionized cloudlets (Redfield and Linsky 2008), it would appear that there are sufficient interfaces for O VI to form if the hot million degree gas is ubiquitous throughout the Local Bubble. However, our observations indicate a distinctly patchy spatial distribution for O VI. Thus, it would appear that the hot gas density within the Local Bubble may be generally too low to produce detectable O VI at most cloud interfaces, but that there are regions, probably at the boundaries of the cavity and at interstellar cloud interfaces where there is sufficiently dense (or compressed) hot material. We note that the evolution of a SNR located within an existing low-density interstellar environment (i.e. a similar situation to that of the Local Bubble cavity) has been modeled for emission from high ions (Shelton 1998, 1999). These models predict that a middle-aged SNR should have a low density and hot (million degree) gas in the interior and a zone of O VI-rich gas at the boundary. With the passage of time the million degree gas in the cavity interior cools sufficiently for its X-ray emission to fade significantly (perhaps to the level calculated by Koutroumpa et al 2009). The SNR cooling model predicts that within a timescale of only $\sim 2.5 \times 10^5$ years the transition temperature ions such as N V and O VI are to be mainly found at the periphery of the cooling SNR bubble and the rapidly cooling central gas is over-ionized and may be out of equilibrium. By $10^6$ years (i.e. less than the age of the Local Bubble), the inner cavity regions will have cooled sufficiently that the level of X-ray emission is negligible, while appreciable amounts of O VI ions should lie within the transition temperature gas at the bubble's edge. The data for our mid-plane detections of O VI could fit such a scenario.

Finally, we note the presence of a particularly interesting pair of stars residing in the region of the l=340° slice, where a detection and non-detection are spatially very close together (nos. 32 and 33; WD2004-605, and WD2014-575), lying on very similar lines of sight with the latter being just 5 pc more distance. The detection (WD2004-605) is weak, $\sim 4\sigma$, so it is conceivable that this is a threshold effect where the additional distance to this object allows the O VI column density to exceed the detectable level. However, there are other non-detections at similar or greater distances within the local bubble (e.g. stars 1, 8, 17, 19, 35, 36) along other lines-of-sight.

4.5 *Towards an understanding of the O VI-SXRB connection*

The data presented in Figures 7 and 8 suggest that there is a general deficiency of O VI gas within the Local Cavity. In addition, for local detections of O VI near the galactic plane a sight-line must either cross or be located at the edge of a neutral cloud at the Local Bubble boundary. Plots of the E-vector polarization of starlight by dust in the interstellar medium presented by Mathewson and Ford (1970) show that, for sight-lines < 50pc, optical polarization is minimal in all galactic directions with no preferential polarization vector angle. However, the polarization for stars in the 50 – 100pc range clearly exhibits large features in the longitude range l = 0 to 60° reaching up from the galactic plane to high +ve latitudes. This polarization vector strengthens for stars in the 100 -200pc distance range, and a similar (but weaker) feature also emerges towards negative galactic latitudes. This feature in the optical polarization data (which results from the alignment of interstellar dust and gas due to a magnetic field) is also mimicked in maps of the galactic distribution of polarized 408MHz radio emission (Wolleben 2007). The feature is directly associated with the North Polar Spur (NPS), which is the brightest filament of the extensive Loop I superbubble (Heiles 1998). It is widely believed that the Loop I superbubble was formed by stellar winds and supernovae events caused by the Sco-Cen OB association of young stars. Its emission has been mapped in both X-rays and radio emission, and Egger (1998) argues that Loop I is caused by a shock from a recent supernova that has heated the inner wall of the Sco-Cen supershell. It is highly likely that this event also could have caused a blow-out from the galactic plane in the form of a galactic fountain.

In the various panels of Figure 8 we have shown the spatial distribution of cold and neutral gas in the meridian plane of the galaxy within 200pc, as traced by NaI absorption measurements (Lallement et al 2003) for comparison with our O VI absorption measurements. These maps clearly show the Local Chimney feature (Welsh et al 1999) from several different viewpoints. This is an extension of the Local Cavity reaching towards high and low galactic latitudes. It was first noted by Welsh et al (1999) that the angle of the Local Chimney is tilted towards the directions where the SXRB emission is greatest in both galactic hemispheres. The absorption characteristics of gas lying close to the axis of the Chimney have been probed by Crawford et al (2002) and Welsh, Sallmen & Lallement (2004). These observations show that interstellar gas with velocities in the -20 to -60 km/s range appears to be falling down the Chimney towards the galactic disk. The origin of this in-falling gas was thought to be part of the Intermediate Velocity Arch feature that resides in the overlying galactic halo. However as stated above, the recent work of Wolleben (2007) argues in favor of an interaction between gas ejected from the adjacent interstellar cavity of Loop I and the Local Cavity.

## 5. CONCLUSIONS

We have carried out a comprehensive survey of O VI absorption in the local interstellar medium using the large sample of white dwarf spectra available in the *FUSE* data archive. Since the *FUSE* mission has now ended and in the absence of any plans for a future space facility covering the same wavelength range, this work will provide the most complete analysis for the foreseeable future. In total, this work covers 80 lines of sight, with a sensitivity varying according to the brightness of the background white dwarf and the exposure time allocated. Hence, we have expanded by factors of 3.6 and 1.7 respectively the number of sight lines included in the earlier work of Oegerle et al. (2005) and Savage and Lehner (2006).

We have detected interstellar O VI along 26 lines of sight in this sample. In most of these the O VI velocity is close to that of the neutral ISM, but one star has O VI blue-shifted relative to the ISM while a few others have red-shifted components. There are very few detections of O VI within the inner rarefied regions of the Local Bubble, with the nearest detection of interstellar OVI being towards WD2004-605 at a distance of 58pc. The derived upper limits to detections within the bubble provide tight constraints on the amount of hot gas that might be present.

In support of previous observations, we find that the spatial distribution of O VI is best described as being "patchy" within 200pc. Although O VI has been detected along sight-lines at high galactic latitudes and towards regions of enhanced soft X-ray background flux, there are many more cases where this is not the case. Therefore, we find no strong evidence to support a spatial correlation between O VI and SXRB emission, thus refuting the initial findings of Welsh and Lallement (2008). However, we find that the majority of white dwarfs yielding interstellar detections of O VI are mostly located close to or beyond the neutral boundary to the Local Cavity in regions of higher gas density. Hence, any T ~ 300,000K gas responsible for the O VI absorption may reside at the interface between the Cavity and surrounding medium or in that medium itself.

This work is based on data from the NASA-CNES-CSA *FUSE* mission, operated by Johns Hopkins University. The data were obtained from the MAST archive operated by the Space Telescope Science Institute. MAB and DDB acknowledge the support of the Science and Technology Facilities Council (UK).

APPENDIX
COMMENTS ON INDIVIDUAL STARS AND LINES-OF-SIGHT

In this section we provide individual comments about some of the stars and the interstellar absorption observed. Measurements associated with these objects are summarised in Tables 3, 4 and 5.

*WD0027-636.* – Strong Si IV and P V. The O VI feature is flat-bottomed in this object and spans both photospheric and interstellar velocities. If the feature is purely stellar, it is asymmetric compared to the stellar velocity. Therefore, this appears to have blended, but unresolved, interstellar and photospheric components.

*WD0050-332.* – Although this is a relatively cool DA at 34,684K, photospheric material has been detected in the STIS spectrum (Barstow et al. 2003; Bannister et al. 2003) and P V is present in the *FUSE* spectrum. The O VI is clearly aligned with the stellar velocity.

*WD0131-164.* – Detections of photospheric Si IV and P V (see figure 1). The O VI velocity is aligned with these velocities.

*WD0026-615.* – Detections of photospheric Si IV and P IV. The O VI velocity is aligned with these velocities.

*WD0232+035.* – Feige 24 is a well-studied hot DA+dM binary system known to contain photospheric heavy elements, also detected in the *FUSE* spectrum (e.g. Si IV and P V). The photospheric velocity varies with phase about the ~4.25d period. In this observation the photospheric and interstellar components are not resolved, so the O VI is assumed to be photospheric.

*WD0252-055.* – Detection of photospheric Si IV and P IV. ISM velocity aligned with stellar velocity. Therefore, O VI assumed to be a stellar component.

*WD0353+284.* – O VI component is at a velocity red-shifted when compared to the interstellar velocity.

*WD0455-282.* – A complex line of sight with two interstellar velocity components (see Table 3). There are both photospheric and interstellar O VI components, with the latter spanning the two interstellar lines.

*WD0512+326.* – Detection of photospheric Si IV. The O VI velocity is aligned with these velocities.

*WD0603-483.* – There are two interstellar velocity components as noted in Table 3, clearly detected in C II, but also weakly seen in O I.

*WD0621-376.* – Strong detections of photospheric material in both *IUE* (e.g. Holberg et al. 1998) and *FUSE*. The O VI is aligned with the photospheric velocity although the photospheric and interstellar components are not clearly resolved.

*WD0715-704.* – The O VI velocity is aligned with the interstellar velocity. No photospheric material is detected in this star.

*WD0802+413.* – Two interstellar components are detected in this star, as detailed in Table 3, but the weaker (at ~70 km s$^{-1}$) is seen only in C II.

*WD1021+266.* – Two interstellar components are detected in this star, as detailed in Table 3, but the weaker (at ~50 km s$^{-1}$) is seen only in C II. Photospheric C III, Si IV and S IV are detected at 13.8 km s$^{-1}$. The O VI velocity is aligned with this velocity.

*WD1029+537.* – Photospheric C III, Si IV, S VI and P V are detected at 12.0 km s$^{-1}$. The O VI velocity is aligned with this velocity.

*WD1040+492.* – Multiple interstellar components detected in both O I and C II. The components are resolved in O I but blended and saturated in C II.

*WD1057+719.* – The O I is contaminated by a geocoronal emission line blueward of the absorption core. The O VI velocity is redward of the ISM value.

*WD1254+223.* – The O VI absorption is board overlapping with the interstellar velocity but redward of with a centroid of 17.4 km s$^{-1}$.

*WD1302+597.* – The O I interstellar line is completely filled in by geocoronal emission, but C II is detected.

*WD1314+293.* – The O VI absorption is board overlapping with the interstellar velocity but redward of with a centroid of 19.8 km s$^{-1}$. This star and WD1254+223 both lie near the north Galactic pole.

*WD1550+130.* – Two ISM components are present. The weaker of the two at 43.8 km s$^{-1}$ is only detected in the C II.

*WD1611-084.* – Photospheric material has been detected in the HST/GHRS spectrum of this star (e.g. Barstow et al. 2003). The photospheric and ISM velocities are not well-separated. Therefore, the O VI is taken to be photospheric. There are two interstellar C II components present.

*WD1620+64.* – The C II interstellar line has a strongly asymmetric profile with and extended red wing, indicating that more than one component may be present. The O I line is filled in by geocoronal emission.

*WD1636+351.* – There are two interstellar components at -31.1 and 24.9 km s$^{-1}$. The weaker redward component is not detected in O I.

*WD1725+586.* – There are at least two strong, saturated components detected in C II, which are barely resolved. No O I is detected as the feature is filled in by strong geocoronal emission. The detected O VI component coincides with the redward ISM line and there may be a weaker O VI line aligned with the blueward ISM component. However, this falls below our detection threshold. The 1037Å O VI line is also detected in this star with a measured column density of 7.3x10$^{-13}$ cm$^{-2}$, almost twice that of the 1031Å line. This suggests that the 1031 line may be saturated. Since the velocities of the two lines are in very close agreement, this could be a rare example of unresolved interstellar O VI tracing a temperature far out of equilibrium.

*WD1845+683.* – Both C II and O I are detected but the O I is contaminated by a geocoronal line.

*WD1942+499.* – Si IV and P V are detected in this star and the O VI detected is clearly aligned with this velocity component rather than the C II and O I.

*WD1950-432.* – While photospheric material (P IV, Si IV, S IV) is identified in this star, the O VI velocity is well-aligned with the interstellar velocity.

*WD2000-561.* – Stellar Si IV and P IV are detected. The measured O VI velocity lies between the interstellar and photospheric velocities. Therefore, the O VI could be photospheric. The interstellar O I is contaminated by a geocoronal contribution.

*WD2011+398.* – Stellar S IV and P IV are detected. The measured O VI velocity aligns with the photospheric value.

*WD2146-433.* – Stellar S IV and P IV are detected. The measured O VI velocity aligns with the photospheric value.

*WD2321-549, WD2331-475, WD2350-706.* – The O VI absorption feature is very broad in all three of these stars, spans both interstellar and photospheric velocities and appears to be a multiple, although not resolved component.

TABLE 1
SUMMARY OF WHITE DWARFS TARGETS AND THEIR PHYSICAL PARAMETERS.

| WD No. | Alt. Name/Cat* | l(°) | b(°) | d(pc) | Reference | Teff | log g | Reference |
|---|---|---|---|---|---|---|---|---|
| WD0001+433 | REJ, EUVEJ | 113.90 | -18.44 | 99 | This work | 46205 | 8.85 | Marsh et al. 1997 |
| WD0004+330 | GD2 | 112.48 | -28.69 | 99 | Vennes et al. 1997 | 47936 | 7.77 | Marsh et al. 1997 |
| WD0027-636 | REJ, EUVEJ | 306.98 | -53.55 | 236 | Vennes et al. 1997 | 60595 | 7.97 | Marsh et al. 1997 |
| WD0050-332 | GD659 | 299.14 | -84.12 | 58 | Holberg et al. 1998 | 34684 | 7.89 | Marsh et al. 1997 |
| WD0106-358 | GD683 | 280.88 | -80.81 | 86 | This work | 28580 | 7.90 | Marsh et al. 1997 |
| WD0131-164 | GD984 | 167.26 | -75.15 | 96 | Vennes et al. 1997 | 44850 | 7.96 | Marsh et al. 1997 |
| WD0147+674 | REJ, EUVEJ | 128.58 | 5.44 | 99 | Dupuis et al. 2006 | 30120 | 7.70 | Marsh et al. 1997 |
| WD0226-615 | HD15638 | 284.20 | -52.16 | 199 | ESA 1997 | 50000 | 8.15 | Landsman et al. 1993 |
| WD0229-481 | REJ, EUVEJ | 266.62 | -61.59 | 240 | This work | 63400 | 7.43 | Marsh et al. 1997 |
| WD0232+035 | Feige 24 | 165.97 | -50.27 | 74 | ESA 1997 | 62947 | 7.53 | Marsh et al. 1997 |
| WD0236+498 | REJ, EUVEJ | 140.15 | -9.15 | 96 | Vennes et al. 1997 | 33822 | 8.47 | Finley et al. 1997 |
| WD0235-125 | PHL1400 | 165.97 | -50.27 | 86 | This work | 32306 | 8.44 | Finley et al. 1997 |
| WD0252-055 | HD18131 | 181.86 | -53.47 | 104 | ESA 1997 | 29120 | 7.50 | Vennes et al. 1997 |
| WD0320-539 | LB1663 | 267.30 | -51.64 | 124 | This work | 32860 | 7.66 | Marsh et al. 1997 |
| WD0325-857 | REJ, EUVEJ | 299.86 | -30.68 | 35 | Barstow et al. 1995 | - | - | - |
| WD0346-011 | GD50 | 188.95 | -40.10 | 29 | Vennes et al. 1997 | 42373 | 9.00 | Marsh et al. 1997 |
| WD0353+284 | REJ, EUVEJ | 165.08 | -18.67 | 107 | Burleigh et al. 1997 | 32984 | 7.87 | This work |
| WD0354-368 |  | 238.64 | -49.98 | 400 | Christian et a. 1996 | 53000 | 8.00 | Christian et al. 1996 |
| WD0455-282 | REJ, EUVEJ | 229.29 | -36.17 | 102 | Oegerle et al. 2005 | 58080 | 7.90 | Marsh et al. 1997 |
| WD0501+524 | G191-B2B | 155.95 | 7.10 | 59 | Vennes et al. 1997 | 57340 | 7.48 | Marsh et al. 1997 |
| WD0512+326 | HD33959C | 173.30 | -3.36 | 25 | ESA 1997 | 22750 | 8.01 | Holberg et al.1999 |
| WD0549+158 | GD71 | 192.03 | -5.34 | 49 | Vennes et al. 1997 | 32780 | 7.83 | Barstow et al. 2003 |
| WD0603-483 | REJ, EUVEJ | 255.78 | -27.36 | 178 | Vennes et al. 1997 | 33040 | 7.80 | Marsh et al. 1997 |
| WD0621-376 | REJ, EUVEJ | 245.41 | -21.43 | 78 | Holberg et al. 1998 | 62280 | 7.22 | Marsh et al. 1997 |
| WD0659+130 | REJ0702+129 | 202.51 | 8.19 | 115 | Vennes et al. 1998 | 39960 | 8.31 | This work |
| WD0715-704 | REJ, EUVEJ | 281.40 | -23.50 | 94 | Vennes et al. 1997 | 44300 | 7.69 | Marsh et al. 1997 |
| WD0802+413 | KUV | 179.22 | 30.94 | 230 | Dupuis et al. 2006 | 45394 | 7.39 | Finley et al. 1997 |
| WD0830-535 | REJ, EUVEJ | 270.11 | -8.27 | 82 | Vennes et al. 1997 | 29330 | 7.79 | Marsh et al. 1997 |
| WD0937+505 | PG | 166.90 | 47.12 | 218 | Dupuis et al. 2006 | 35552 | 7.76 | Finley et al. 1997 |
| WD1019-141 | REJ, EUVEJ | 256.48 | 34.74 | 112 | This work | 29330 | 7.79 | Marsh et al. 1997 |
| WD1021+266 | HD90052 | 205.72 | 57.21 | 250 | Burleigh et al. 1997 | 35432 | 7.48 | This work |
| WD1024+326 | REJ, EUVEJ | 194.52 | 58.41 | 400 | This work | 41354 | 7.59 | This work |
| WD1029+537 | REJ, EUVEJ | 157.51 | 53.24 | 116 | Vennes et al. 1997 | 44980 | 7.68 | Marsh et al. 1997 |
| WD1040+492 | REJ, EUVEJ | 162.67 | 57.01 | 230 | Vennes et al. 1997 | 47560 | 7.62 | Marsh et al. 1997 |
| WD1041+580 | REJ, EUVEJ | 150.12 | 52.17 | 93 | Vennes et al. 1997 | 29016 | 7.79 | Marsh et al. 1997 |
| WD1057+719 | PG | 134.48 | 42.92 | 141 | Savage & Lehner 2006 | 39555 | 7.66 | Marsh et al. 1997 |
| WD1109-225 | HD97277 | 274.78 | 34.54 | 82 | ESA 1997 | 36750 | 7.50 | Barstow et al.1994 |
| WD1234+481 | PG | 129.81 | 69.01 | 129 | Vennes et al. 1997 | 55570 | 7.57 | Marsh et al. 1997 |
| WD1254+223 | GD153 | 317.25 | 84.75 | 67 | Vennes et al. 1997 | 38205 | 7.90 | Barstow et al. 2003 |
| WD1302+597 | GD323 | 119.82 | 57.59 | 79 | This work | 39960 | 8.31 | This work |
| WD1314+293 | HZ43 | 54.11 | 84.16 | 68 | Finley et al. 1997 | 49435 | 7.95 | Barstow et al. 2003 |
| WD1337+701 | EG102 | 117.18 | 46.31 | 104 | Dupuis et al. 2006 | 20435 | 7.87 | Finley et al. 1997 |
| WD1342+442 | PG | 94.30 | 69.92 | 387 | This work | 66750 | 7.93 | Barstow et al. 2003 |
| WD1440+753 | REJ, EUVEJ | 114.10 | 40.12 | 98 | Vennes et al. 1997 | 42400 | 8.54 | Vennes et al. 1997 |
| WD1528+487 | REJ, EUVEJ | 78.87 | 52.72 | 140 | Vennes et al. 1997 | 46230 | 7.70 | Marsh et al. 1997 |
| WD1550+130 | NN Ser | 23.65 | 45.34 | 756 | This work | 39910 | 6.82 | This work |
| WD1603+432 | PG | 68.22 | 47.93 | 114 | Dupuis et al. 2006 | 36257 | 7.85 | Finley et al. 1997 |
| WD1611-084 | REJ, EUVEJ | 4.30 | 29.30 | 93 | Vennes 1999 | 38500 | 7.85 | Marsh et al. 1997 |
| WD1615-154 | G153-41 | 358.79 | 24.18 | 55 | Oegerle et al. 2005 | 38205 | 7.90 | Barstow et al. 2003 |
| WD1620+647 | PG | 96.61 | 40.16 | 174 | Dupuis et al. 2006 | 30184 | 7.72 | Finley et al. 1997 |
| WD1631+781 | REJ, EUVEJ | 111.30 | 33.58 | 67 | Vennes et al. 1997 | 44559 | 7.79 | Finley et al. 1997 |
| WD1635+529 | HD150100 | 80.94 | 41.49 | 123 | ESA 1997 | 20027 | 8.14 | This work |
| WD1636+351 | WD1638+349 | 56.98 | 41.40 | 109 | Vennes et al. 1997 | 36056 | 7.71 | Marsh et al. 1997 |
| WD1648+407 | REJ, EUVEJ | 64.64 | 39.60 | 200 | Vennes et al. 1997 | 37850 | 7.95 | Marsh et al. 1997 |
| WD1725+586 | LB335 | 87.17 | 33.83 | 123 | This work | 54550 | 8.49 | Marsh et al. 1997 |

| WD No. | Alt. Name/Cat* | l(°) | b(°) | d(pc) | Reference | Teff | log g | Reference |
|---|---|---|---|---|---|---|---|---|
| WD1800+685 | KUV | 98.73 | 29.78 | 159 | Vennes et al. 1997 | 43701 | 7.80 | Marsh et al. 1997 |
| WD1819+580 | REJ, EUVEJ | 87.00 | 26.76 | 103 | This work | 45330 | 7.73 | Marsh et al. 1997 |
| WD1844-223 | REJ, EUVEJ | 12.50 | -9.25 | 62 | Vennes et al. 1997 | 31470 | 8.17 | Finley et al. 1997 |
| WD1845+683 | KUV | 98.81 | 25.65 | 125 | Vennes et al. 1997 | 36888 | 8.12 | Finley et al. 1997 |
| WD1917+509 | REJ, EUVEJ | 91.02 | 20.04 | 105 | Vennes et al. 1997 | 33000 | 7.90 | Vennes et al. 1997 |
| WD1921-566 | REJ, EUVEJ | 340.63 | -27.01 | 110 | Vennes et al. 1998 | 52946 | 8.16 | This work |
| WD1942+499 |  | 83.08 | 12.75 | 105 | Vennes et al. 1997 | 33500 | 7.86 | Marsh et al. 1997 |
| WD1950-432 | HS | 356.51 | -29.09 | 140 | Dupuis et al. 2006 | 41339 | 7.85 | Finley et al. 1997 |
| WD2000-561 | REJ, EUVEJ | 341.78 | -32.25 | 198 | Dupuis et al. 2006 | 44456 | 7.54 | Marsh et al. 1997 |
| WD2004-605 | REJ, EUVEJ | 336.58 | -32.86 | 58 | Vennes et al. 1997 | 44200 | 8.14 | Marsh et al. 1997 |
| WD2011+398 | REJ, EUVEJ | 77.00 | 3.18 | 141 | Vennes et al. 1997 | 47057 | 7.74 | Marsh et al. 1997 |
| WD2014-575 | LS210-114 | 340.20 | -34.25 | 51 | Vennes et al. 1997 | 26579 | 7.78 | Marsh et al. 1997 |
| WD2020-425 | REJ, EUVEJ | 358.36 | -34.45 | 52 | This work | 28597 | 8.54 | Marsh et al. 1997 |
| WD2111+498 | GD394 | 91.37 | 1.13 | 50 | Vennes et al. 1997 | 38866 | 7.84 | Marsh et al. 1997 |
| WD2116+736 | KUV | 109.39 | 16.92 | 177 | Dupuis et al. 2006 | 54486 | 7.76 | Finley et al. 1997 |
| WD2124+191 | HD204188 | 70.43 | -21.98 | 46 | Vennes 1998 | 35000 | 9.00 | Landsman et al. 1993 |
| WD2124-224 | REJ, EUVEJ | 27.36 | -43.76 | 224 | Oegerle et al. 2005 | 48297 | 7.69 | Marsh et al. 1997 |
| WD2146-433 | REJ, EUVEJ | 357.18 | -50.13 | 362 | Dupuis et al. 2006 | 67912 | 7.58 | Finley et al. 1997 |
| WD2152-548 | REJ, EUVEJ | 340.50 | -48.70 | 129 | - | 45800 | 7.78 | Vennes et al. 1997 |
| WD2211-495 | REJ, EUVEJ | 345.79 | -52.62 | 53 | Vennes et al. 1997 | 65600 | 7.42 | Marsh et al. 1997 |
| WD2257-073 | HD217411 | 65.17 | -56.93 | 89 | Vennes et al. 1998 | 38010 | 7.84 | This work |
| WD2309+105 | GD246 | 87.26 | -45.12 | 79 | Vennes et al. 1997 | 51300 | 7.91 | Barstow et al. 2003 |
| WD2321-549 | REJ, EUVEJ | 326.91 | -58.21 | 192 | Dupuis et al. 2006 | 45860 | 7.73 | Marsh et al. 1997 |
| WD2331-475 | REJ, EUVEJ | 334.84 | -64.81 | 82 | Vennes et al. 1997 | 56682 | 7.64 | Marsh et al. 1997 |
| WD2350-706 | HD223816 | 309.91 | -45.94 | 92 | Barstow et al. 2001 | 69300 | 8.00 | Barstow et al. 1994 |

\* PG = Palomar Green, HS = Hamburg Schmidt, REJ = Rosat EUV, EUVEJ = EUVE, KUV = Kiso UV, KPD = Kitt Peak Downes

TABLE 2
SUMMARY OF *FUSE* OBSERVATIONS

| WD No. | Alt. Name/Cat | Observation | Start Time |
|---|---|---|---|
| WD0001+433 | REJ, EUVEJ | E90305010 | 2004-11-24 19:34:00 |
| WD0004+330 | GD2 | P20411020-40 | 2002-09-14 10:07:00 |
| WD0027-636 | REJ, EUVEJ | Z90302010 | 2002-09-30 15:44:00 |
| WD0050-332 | GD659 | M10101010 | 2000-07-04 23:52:00 |
|  |  | P20420010 | 2000-12-11 07:00:00 |
| WD0106-358 | GD683 | D02301010 | 2004-11-21 06:29:00 |
| WD0111+002 |  | A01303030 | 2001-08-07 12:43:00 |
| WD0131-164 | GD984 | P20412010 | 2000-12-10 07:45:00 |
| WD0147+674 | REJ, EUVEJ | Z90303020 | 2002-10-12 17:08:00 |
| WD0226-615 | HD15638 | A05402010 | 2000-07-04 20:39:00 |
| WD0229-481 | REJ, EUVEJ | M10504010 | 2002-09-21 11:51:00 |
| WD0232+035 | Feige 24 | P10405040 | 2004-01-06 01:00:00 |
| WD0236+498 | REJ, EUVEJ | Z90306010 | 2002-12-11 16:47:00 |
| WD0235-125 | PHL1400 | E56809010 | 2004-08-30 17:27:00 |
| WD0252-055 | HD18131 | A05404040 | 2001-11-28 06:39:00 |
| WD0320-539 | LB1663 | D02302010 | 2003-09-09 08:50:00 |
| WD0325-857 | REJ, EUVEJ | A01001010 | 2000-09-17 19:52:00 |
| WD0346-011 | GD50 | B12201020-50 | 2003-12-20 04:43:00 |
| WD0353+284 | REJ, EUVEJ | B05510010 | 2001-01-02 22:31:00 |
| WD0354-368 |  | B05511010 | 2001-08-12 11:18:00 |
| WD0416+402 | KPD | Z90308010 | 2003-10-05 13:39:00 |
| WD0455-282 | REJ, EUVEJ | P10411030 | 2000-02-07 09:36:00 |
| WD0501+524 | G191-B2B | Coadded* |  |
| WD0512+326 | HD33959C | A05407070 | 2000-03-01 04:08:00 |
| WD0549+158 | GD71 | P20417010 | 2000-11-04 15:32:00 |
| WD0603-483 | REJ, EUVEJ | Z90309010 | 2002-12-31 19:26:00 |
| WD0621-376 | REJ, EUVEJ | P10415010 | 2000-12-06 05:28:00 |
| WD0659+130 | REJ0702+129 | B05509010 | 2001-03-05 21:53:00 |
| WD0715-704 | REJ, EUVEJ | M10507010 | 2003-08-15 00:08:00 |
| WD0802+413 | KUV | Z90311010 | 2004-03-14 09:37:00 |
| WD0830-535 | REJ, EUVEJ | Z90313010 | 2002-05-03 13:37:00 |
| WD0937+505 | PG | Z90314010 | 2003-03-24 09:13:00 |
| WD1019-141 | REJ, EUVEJ | P20415010 | 2001-05-15 21:35:00 |
| WD1021+266 | HD90052 | B05508010 | 2001-05-01 19:34:00 |
| WD1024+326 | REJ, EUVEJ | B05512010 | 2001-05-01 09:22:00 |
| WD1029+537 | REJ, EUVEJ | B00301010 | 2001-03-25 04:40:00 |
| WD1040+492 | REJ, EUVEJ | Z00401010 | 2002-04-04 12:42:00 |
| WD1041+580 | REJ, EUVEJ | Z90317010 | 2002-04-07 10:04:00 |
| WD1057+719 | PG | Z90318010 | 2002-04-08 16:12:00 |
| WD1109-225 | HD97277 | A05401010 | 2000-05-29 16:00:00 |
| WD1234+481 | PG | M10524020 | 2003-03-18 20:29:00 |
| WD1254+223 | GD153 | M10104020 | 2000-04-29 21:32:00 |

TABLE 2: continued

| | | | |
|---|---|---|---|
| WD1302+597 | GD323 | C02601010 | 2002-04-02 17:00:00 |
| WD1314+293 | HZ43 | P10423010 | 2000-04-22 21:19:00 |
| WD1337+701 | EG102 | B11903010 | 2001-05-05 06:43:00 |
| WD1342+442 | PG | A03404020 | 2000-01-11 23:01:00 |
| WD1440+753 | REJ, EUVEJ | Z90322010 | 2003-01-15 08:40:00 |
| WD1528+487 | REJ, EUVEJ | P20401010 | 2001-03-27 23:14:00 |
| WD1550+130 | NN Ser | E11201010 | 2004-04-15 05:38:00 |
| WD1603+432 | PG | Z90324010 | 2003-07-02 11:51:00 |
| WD1611-084 | REJ, EUVEJ | B11904010 | 2004-04-02 11:53:00 |
| WD1615-154 | G153-41 | P20419010 | 2001-08-29 12:40:00 |
| WD1620+647 | PG | Z90325010 | 2002-05-17 04:45:00 |
| WD1631+781 | REJ, EUVEJ | M10528040 | 2004-02-13 20:02:00 |
| WD1635+529 | HD150100 | C05002010 | 2002-07-13 14:00:00 |
| WD1636+351 | WD1638+349 | P20402010 | 2001-03-28 14:22:00 |
| WD1648+407 | REJ, EUVEJ | Z90326010 | 2002-07-12 11:22:00 |
| WD1725+586 | LB335 | Z90327010 | 2002-05-13 23:16:00 |
| WD1800+685 | KUV | M10530010-70 | 2002-09-09 15:14:53 |
| | | P20410010-20 | 2001-10-01 04:55:36 |
| WD1819+580 | REJ, EUVEJ | Z90328010 | 2002-05-11 01:25:00 |
| WD1844-223 | REJ, EUVEJ | P20405010 | 2001-04-28 20:01:00 |
| WD1845+683 | KUV | Z90329010 | 2002-05-08 20:40:00 |
| | | Z99003010 | 2002-09-13 10:39:00 |
| WD1917+509 | REJ, EUVEJ | Z90330010 | 2002-05-07 01:36:00 |
| WD1921-566 | REJ, EUVEJ | A05411110 | 2000-05-24 22:59:00 |
| WD1942+499 | | Z90331010 | 2002-05-08 00:12:00 |
| WD1950-432 | HS | Z90332010 | 2002-10-04 09:18:00 |
| WD2000-561 | REJ, EUVEJ | Z90333010 | 2002-04-15 21:30:00 |
| WD2004-605 | REJ, EUVEJ | P20422010 | 2001-05-21 21:09:00 |
| WD2011+398 | REJ, EUVEJ | M10531020 | 2003-10-23 12:19:00 |
| WD2014-575 | LS210-114 | Z90334010 | 2002-04-16 15:55:00 |
| WD2020-425 | REJ, EUVEJ | Z90335010 | 2002-06-07 09:16:00 |
| WD2043-635 | BPM13537 | Z90337010 | 2002-04-13 09:10:00 |
| WD2111+498 | GD394 | M10532010 | 2002-10-27 16:44:00 |
| WD2116+736 | KUV | Z90338020 | 2002-10-28 01:26:00 |
| WD2124+191 | HD204188 | A05409090 | 2001-07-12 22:23:00 |
| WD2124-224 | REJ, EUVEJ | P20406010 | 2001-10-08 20:26:00 |
| WD2146-433 | REJ, EUVEJ | Z90339010 | 2002-10-06 15:14:00 |
| WD2152-548 | REJ, EUVEJ | M10515010 | 2002-09-24 03:07:00 |
| WD2211-495 | REJ, EUVEJ | M10303160 | 2003-05-24 23:28:00 |
| WD2257-073 | HD217411 | A05410100 | 2000-06-28 11:05:00 |
| WD2309+105 | GD246 | P20424010 | 2001-07-14 01:26:00 |
| WD2321-549 | REJ, EUVEJ | Z90342010 | 2002-07-22 05:10:00 |
| WD2331-475 | REJ, EUVEJ | M10517010 | 2002-09-23 08:46:00 |
| WD2350-706 | HD223816 | A05408090 | 2001-05-23 16:21:00 |

TABLE 2: continued

| | | |
|---|---|---|
| *G191-B2B obs. | M1010201000 | 1999-10-13 01:25:00 |
| | M1030501000 | 1999-11-12 07:35:00 |
| | M1030502000 | 1999-11-20 07:22:00 |
| | M1030401000 | 1999-11-20 09:02:00 |
| | M1030603000 | 1999-11-20 10:43:00 |
| | M1030503000 | 1999-11-21 06:43:00 |
| | M1030504000 | 1999-11-21 10:03:00 |
| | M1030602000 | 1999-11-21 11:39:00 |
| | S3070101000 | 2000-01-14 09:40:00 |
| | M1010202000 | 2000-02-17 06:10:00 |
| | M1030604000 | 2001-01-09 09:02:00 |
| | M1030506000 | 2001-01-09 09:26:00 |
| | M1030605000 | 2001-01-10 13:20:00 |
| | M1030507000 | 2001-01-10 13:45:00 |
| | M1030403000 | 2001-01-10 15:08:00 |
| | M1030606000 | 2001-01-23 06:08:00 |
| | M1030508000 | 2001-01-23 07:55:00 |
| | M1030404000 | 2001-01-23 11:18:00 |
| | M1030607000 | 2001-01-25 04:46:00 |
| | M1030509000 | 2001-01-25 06:33:00 |
| | M1030405000 | 2001-01-25 09:53:00 |
| | M1030608000 | 2001-09-28 13:50:00 |
| | M1030510000 | 2001-09-28 15:35:00 |
| | M1030406000 | 2001-09-28 17:15:00 |
| | M1030609000 | 2001-11-21 09:54:00 |
| | M1030511000 | 2001-11-21 11:39:00 |
| | M1030407000 | 2001-11-21 13:19:00 |
| | M1030610000 | 2002-02-17 07:27:00 |
| | M1030512000 | 2002-02-17 12:34:00 |
| | M1030408000 | 2002-02-17 17:43:00 |
| | M1030611000 | 2002-02-23 02:05:00 |
| | M1030513000 | 2002-02-23 06:43:00 |
| | M1030409000 | 2002-02-23 08:23:00 |
| | M1030612000 | 2002-02-25 02:17:00 |
| | M1030514000 | 2002-02-25 06:59:00 |
| | M1030613000 | 2002-12-03 21:00:00 |
| | M1030614000 | 2002-12-06 02:30:00 |
| | M1030515000 | 2002-12-06 05:16:00 |
| | M1052001000 | 2002-12-07 21:46:00 |
| | M1030615000 | 2002-12-08 22:36:00 |
| | M1030516000 | 2002-12-09 00:29:00 |
| | M1030412000 | 2002-12-09 03:53:00 |
| | M1030616000 | 2003-02-05 19:14:00 |

TABLE 2: continued

| | |
|---|---|
| M1030517000 | 2003-02-05 21:07:00 |
| M1030413000 | 2003-02-06 00:36:00 |
| S4057604000 | 2003-11-21 05:21:00 |
| M1030617000 | 2003-11-23 20:16:00 |
| M1030519000 | 2004-01-25 21:31:00 |
| M1030415000 | 2004-01-26 00:51:00 |

TABLE 3
MEASURED PHOTOSPHERIC AND INTERSTELLAR VELOCITIES (KM S$^{-1}$) FOR EACH WHITE DWARF WHERE O VI 1031.912Å IS DETECTED.

| WD No. | $v_{phot}$ | $v_{ism}$ | δv | $v_{OVI}$ | Ew | σ | ew/σ | $|v_{OVI}-v_{phot}|$ | $|v_{OVI}-v_{ism}|$ | O VI? | S&L |
|---|---|---|---|---|---|---|---|---|---|---|---|
| WD0001+433 | 7.0 | -10.7 | 17.7 | | <23.4 | | | | | ND | |
| WD0004+330 | | -1.8 | | -11.0 | 6.1 | 1.2 | 5.0 | 11.0 | 9.2 | I | I |
| WD0027-636 | 21.8 | -5.5 | 27.3 | 0.5 | 28.6 | 4.6 | 6.2 | 21.3 | 6.0 | I+P | P |
| WD0050-332 | 18.4 | -14.9 | 33.4 | 13.7 | 6.3 | 1.7 | 3.8 | 4.8 | 28.6 | P | P |
| WD0106-358 | -12.1 | -5.0 | -7.1 | | <15.3 | | | | | ND | |
| WD0131-164 | 10.9 | -8.6 | 19.5 | 19.5 | 43.2 | 3.0 | 14.5 | 8.6 | 28.1 | P | |
| WD0147+674 | | -15.5 | | | <26.1 | | | | | ND | ND |
| WD0226-615 | 25.8 | -5.5 | 31.2 | 9.3 | 16.4 | 4.5 | 3.6 | 16.5 | 14.8 | P | |
| WD0229-481 | 23.0 | 0.9 | 22.1 | | <12.0 | | | | | ND | |
| WD0232+035 | 4.0 | -7.6 | 11.5 | -3.5 | 8.4 | 1.4 | 5.9 | 7.5 | 4.1 | P | |
| WD0235-125 | | 13.7 | | | <20.6 | | | | | ND | |
| WD0236+498 | | -4.5 | | | <19.8 | | | | | ND | |
| WD0252-055 | 10.4 | 12.2 | -1.8 | 16.2 | 23.9 | 6.7 | 3.6 | 5.8 | 4.0 | P | |
| WD0320-539 | | 2.5 | | | <15.8 | | | | | ND | |
| WD0325-857 | | -10.0 | | | <5.3 | | | | | ND | |
| WD0346-011 | | 15.8 | | | <5.7 | | | | | ND | |
| WD0353+284 | | 9.0 | | 25.6 | 69.9 | 11.3 | 6.2 | | 16.6 | IR | |
| WD0354-368 | | 8.7 | | | <44.9 | | | | | ND | |
| WD0455-282 | 57.7 | -14.2/-65.5 | 71.9 | 59.3 | 11.3 | 2.3 | 5.0 | 1.6 | 73.4 | IB+P | I+P |
| WD0501+524 | 21.5 | 14.3 | 7.2 | 20.4 | 4.8 | 0.5 | 10.7 | 1.1 | 6.1 | P | P |
| WD0512+326 | 20.5 | -3.9 | 24.4 | 25.0 | 14.4 | 4.4 | 3.3 | 4.4 | 28.8 | P | |
| WD0549+158 | 8.3 | 1.4 | 6.8 | | <4.0 | | | | | ND | ND |
| WD0603-483 | | 17.5/-34,6 | | | <16.5 | | | | | ND | I* |
| WD0621-376 | 31.3 | 13.2 | 18.1 | 34.3 | 4.7 | 1.0 | 4.5 | 3.0 | 21.1 | P | |
| WD0659+130 | | -0.3 | | | <26.3 | | | | | ND | |
| WD0715-704 | | -4.8 | | -11.3 | 11.6 | 3.1 | 3.7 | | 6.5 | I | I |
| WD0802+413 | 6.9 | 13.9/61.4 | -6.9 | | <33.7 | | | | | ND | ND |
| WD0830-535 | | 11.1 | | -1.9 | 31.0 | 9.0 | 3.4 | | 13.0 | I | I |
| WD0937+505 | | -6.5 | | | 40.7 | 10.7 | 3.8 | | | I | I |
| WD1019-141 | 11.5 | -6.8 | 18.3 | | <10.9 | | | | | ND | I* |
| WD1021+266 | 12.0 | -5.3 | 17.2 | 13.4 | 36.4 | 5.6 | 6.5 | 1.4 | 18.6 | P | |
| WD1024+326 | | -14.0 | | | <51.8 | | | | | ND | |
| WD1029+537 | 13.8 | -22.8 | 36.6 | 25.7 | 63.6 | 4.4 | 14.6 | 11.9 | 48.5 | P | |
| WD1040+492 | -13.9 | -12.6/-47.9 | -1.3 | 45.0 | 30.3 | 3.7 | 8.2 | 58.9 | 57.6 | IR | |
| WD1041+580 | | -10.4 | | | <14.6 | | | | | ND | ND |
| WD1057+719 | | -16.1 | | 19.8 | 26.3 | 2.9 | 9.1 | | 35.9 | IR | I |
| WD1109-225 | | -4.9 | | -8.0 | 13.8 | 4.0 | 3.5 | | 3.1 | I | |
| WD1234+481 | | -9.6 | | -6.4 | 10.4 | 3.0 | 3.5 | | 3.2 | I | I |
| WD1254+223 | | -11.8 | | 17.4 | 15.1 | 5.0 | 3.0 | | 29.2 | IR | I |
| WD1302+597 | | -6.7 | | | <17.4 | | | | | ND | |
| WD1314+293 | | -10.1 | | 19.8 | 10.5 | 1.8 | 5.8 | | 29.8 | IR | I |
| WD1337+701 | -17.2 | -12.1 | -5.2 | | <113.4 | | | | | ND | ND |
| WD1342+442 | -13.1 | -25.6 | 12.5 | -12.2 | 72.3 | 8.0 | 9.1 | 0.9 | 13.3 | P | |
| WD1440+753 | | -14.0 | | | <11.9 | | | | | ND | ND |
| WD1528+487 | | -27.3 | | -34.3 | 23.8 | 4.1 | 5.8 | | 7.0 | I | I |
| WD1550+130 | | -9.9/43.8 | | | <23.0 | | | | | ND | |
| WD1603+432 | | -34.3 | | | <16.1 | | | | | ND | I |
| WD1611-084 | -35.3 | -21.0/40.6 | -14.3 | -24.2 | 53.1 | 4.0 | 13.3 | 11.1 | 3.2 | P | |
| WD1615-154 | | -68.3 | | | <12.0 | | | | | ND | ND |
| WD1620+647 | | -31.7 | | | <17.2 | | | | | ND | ND |
| WD1631+781 | | -11.8 | | -8.0 | 6.9 | 2.1 | 3.3 | | 3.8 | I | I |

TABLE 3 - continued

| WD No. | $v_{phot}$ | $v_{ism}$ | $\delta v$ | $v_{OVI}$ | Ew | $\sigma$ | ew/$\sigma$ | $|v_{OVI}-v_{phot}|$ | $|v_{OVI}-v_{ism}|$ | O VI? | S&L |
|---|---|---|---|---|---|---|---|---|---|---|---|
| WD1635+529 | | -22.5 | | | <16.3 | | | | | ND | |
| WD1636+351 | | -31.1/24.9 | | | <15.1 | | | | | ND | I* |
| WD1648+407 | | -28.4 | | -18.0 | 21.9 | 9.3 | 2.4 | | 10.4 | I | I |
| WD1725+586 | | -37.8/25.4 | | 18.7 | 53.1 | 5.2 | 10.2 | | 56.5 | IR | |
| WD1800+685 | | -29.0 | | -18.9 | 11.7 | 2.1 | 5.5 | | | I | I |
| WD1819+580 | -8.4 | -24.8 | 16.4 | | <13.7 | | | | | ND | |
| WD1844-223 | | -35.5 | | | <7.4 | | | | | ND | ND |
| WD1845+683 | | -28.2 | | | <32.2 | | | | | ND | ND |
| WD1917+509 | 3.1 | -29.3 | 32.4 | | <12.5 | | | | | ND | ND |
| WD1921-566 | | -30.1 | | | <22.2 | | | | | ND | |
| WD1942+499 | 7.5 | -27.8 | 35.3 | 6.4 | 15.7 | 3.1 | 5.1 | 1.1 | 34.2 | P | P |
| WD1950-432 | -26.0 | -8.4 | -17.6 | 0.3 | 17.5 | 5.0 | 3.5 | 26.3 | 8.7 | I | I |
| WD2000-561 | -9.6 | -26.2 | 16.7 | -15.3 | 25.4 | 6.9 | 3.7 | 5.7 | 11.0 | P | P |
| WD2004-605 | | -27.2 | | -18.0 | 9.4 | 2.8 | 3.4 | | 9.2 | I | I |
| WD2011+398 | 11.1 | -12.3 | 23.4 | 13.9 | 62.6 | 3.5 | 18.0 | 2.7 | 26.1 | P | |
| WD2014-575 | | -29.8 | | | <29.0 | | | | | ND | ND |
| WD2020-425 | | -25.4 | | | <42.0 | | | | | ND | |
| WD2111+498 | 24.7 | -20.0 | 44.7 | 13.1 | 20.1 | 4.2 | 4.8 | 11.7 | 33.1 | P | P |
| WD2116+736 | | -17.6 | | -3.5 | 9.9 | 2.6 | 3.8 | | 14.1 | I | I |
| WD2124+191 | | -19.7 | | | <44.7 | | | | | ND | |
| WD2124-224 | 29.7 | -28.4 | 58.1 | -16.4 | 20.1 | 4.7 | 4.3 | 46.1 | 12.0 | I | I |
| WD2146-433 | 27.9 | -13.1 | 41.0 | 10.4 | 20.3 | 4.5 | 4.6 | 17.5 | 23.5 | P | P |
| WD2152-548 | | -3.9 | | -2.6 | 90.0 | 6.6 | 13.7 | | 1.3 | I | |
| WD2211-495 | 23.9 | -8.9 | 32.8 | 18.0 | 15.4 | 1.2 | 13.4 | 5.9 | 26.9 | P | P |
| WD2257-073 | | -10.2 | | 30.8 | 19.1 | 6.2 | 3.1 | | 41.0 | IR | |
| WD2309+105 | -33.1 | -28.7 | -4.4 | | <5.1 | | | | | ND | ND |
| WD2321-549 | 17.5 | -16.6 | 34.2 | -1.6 | 44.5 | 6.3 | 7.0 | 19.1 | 15.0 | I+P | P |
| WD2331-475 | 28.3 | -3.0 | 31.2 | 3.0 | 13.2 | 2.4 | 5.6 | 25.2 | 6.0 | I+P | P |
| WD2350-706 | 40.7 | -6.4 | 47.1 | 10.6 | 45.1 | 3.6 | 12.5 | 30.0 | 17.0 | I+P | |

TABULATED ABOVE ARE THE DIFFERENCE BETWEEN PHOTOSPHERIC AND ISM VELOCITIES; THE WAVELENGTH, VELOCITY, EQUIVALENT WIDTH (MÅ) AND SIGNIFICANCE OF THE O VI 1031.912Å (EW/ UNCERTAINTY IN EW); ABSOLUTE DIFFERENCES BETWEEN THE O VI AND PHOTOSPHERIC/ISM VELOCITIES; INDICATION OF THE NATURE OF THE O VI ABSORPTION (P – PHOTOSPHERIC, I – INTERSTELLAR AT $v_{ISM}$, IR – INTERSTELLAR RED-SHIFTED WRT $v_{ISM}$, IB – INTERSTELLAR BLUE-SHIFTED WRT $v_{ISM}$, ND– NON-DETECTION); NATURE OF THE O VI AS DETERMINED BY SAVAGE & LEHNER (2006). THE DATA ARE GROUPED BY THE NATURE OF THE O VI DETECTED IN THESE STARS. * FOR THESE STARS THE REASON FOR THE DISCREPANCY BETWEEN THIS WORK AND SAVAGE & LEHNER IS THEIR ADOPTION OF A 2 $\sigma$ DETECTION THRESHOLD.

TABLE 4
stars where O VI 1037.613 Å is detected

| WD No. | Star | $v_{OVI (1031.912)}$ | O VI? | $v_{OVI (1037.613)}$ | ew | σ | ew/σ |
|---|---|---|---|---|---|---|---|
| WD0027-636 | REJ, EUVEJ | 0.49 | I+P | 16.79 | 28.07 | 4.53 | 6.20 |
| WD0131-164 | GD984 | 19.46 | P | 18.59 | 50.29 | 2.47 | 20.36 |
| WD0252-055 | HD18131 | 16.24 | P | 27.26 | 28.47 | 4.88 | 5.83 |
| WD0455-282 | REJ, EUVEJ | 59.27 | IB+P | -65.35 | 11.40 | 1.91 | 5.97 |
| WD0512+326 | HD33959C | 24.96 | P | 12.20 | 14.24 | 2.82 | 5.05 |
| WD0621-376 | REJ, EUVEJ | 34.34 | P | 25.15 | 2.90 | 0.85 | 3.41 |
| WD1021+266 | HD90052 | 13.36 | P | 12.27 | 32.08 | 4.74 | 6.77 |
| WD1029+537 | REJ, EUVEJ | 25.68 | P | 24.09 | 50.47 | 3.46 | 14.59 |
| WD1342+442 | PG | -12.23 | P | -20.14 | 69.07 | 6.96 | 9.92 |
| WD1528+487 | REJ, EUVEJ | -34.34 | I | -32.52 | 18.78 | 4.00 | 4.70 |
| WD1611-084 | REJ, EUVEJ | -24.20 | P | -26.24 | 44.96 | 3.93 | 11.44 |
| WD1615-154 | G153-41 | 46.72 | ND | 20.17 | 7.67 | 3.00 | 2.56 |
| WD1725+586 | LB335 | 18.70 | I | 18.09 | 45.75 | 4.94 | 9.26 |
| WD2011+398 | REJ, EUVEJ | 13.89 | P | 11.18 | 44.76 | 2.78 | 16.10 |
| WD2124-224 | REJ, EUVEJ | -16.39 | I | 11.25 | 5.04 | 2.93 | 1.72 |
| WD2152-548 | REJ, EUVEJ | -2.61 | I | -8.99 | 55.42 | 4.78 | 11.59 |
| WD2211-495 | REJ, EUVEJ | 18.04 | P | 18.20 | 13.71 | 1.16 | 11.82 |

TABLE 5

Measured equivalent widths, errors and column and volume densities of interstellar O VI for all the detections reported in this paper.

| WD No. | d(pc) | ew | σ | $N_{OVI}$ (cm$^{-2}$) | Error | $n_{OVI}$ (cm$^{-3}$) | Error |
|---|---|---|---|---|---|---|---|
| WD0001+433 | 99 | 23.4 | | <1.9E+13 | | <4.7E-08 | |
| WD0004+330 | 99 | 6.1 | 1.2 | 4.9E+12 | 9.7E+11 | 1.2E-08 | 2.5E-09 |
| WD0027-636 | 236 | 14.3 | 4.6 | ~1.1E+13 | | ~1.2E-08 | |
| WD0106-358 | 86 | 15.3 | | <1.2E+13 | | <3.6E-08 | |
| WD0147+674 | 99 | 26.1 | | <2.1E+13 | | <5.3E-08 | |
| WD0229-481 | 240 | 12.0 | | <9.6E+12 | | <1.0E-08 | |
| WD0235-125 | 86 | 20.6 | | <1.6E+13 | | <4.8E-08 | |
| WD0236+498 | 96 | 19.8 | | <1.6E+13 | | <4.1E-08 | |
| WD0320-539 | 124 | 15.8 | | <1.3E+13 | | <2.5E-08 | |
| WD0325-857 | 35 | 5.3 | | <4.2E+12 | | <3.0E-08 | |
| WD0346-011 | 29 | 5.7 | | <4.5E+12 | | <3.9E-08 | |
| WD0353+284 | 107 | 69.9 | 11.3 | 5.6E+13 | 9E+12 | 1.3E-07 | 2.1E-08 |
| WD0354-368 | 400 | 44.9 | | <3.6E+13 | | <2.2E-08 | |
| WD0455-282 | 102 | 11.3 | 2.3 | 9.0E+12 | 1.8E+12 | 2.2E-08 | 4.4E-09 |
| WD0549+158 | 49 | 4.0 | | <3.2E+12 | | <1.6E-08 | |
| WD0603-483 | 178 | 16.5 | | <1.3E+13 | | <1.9E-08 | |
| WD0659+130 | 115 | 26.3 | | <2.1E+13 | | <4.6E-08 | |
| WD0715-704 | 94 | 11.6 | 3.1 | 9.2E+12 | 2.5E+12 | 2.5E-08 | 6.7E-09 |
| WD0802+413 | 230 | 33.7 | | <2.7E+13 | | <2.9E-08 | |
| WD0830-535 | 82 | 31.0 | 9.0 | 2.5E+13 | 7.2E+12 | 7.6E-08 | 2.2E-08 |
| WD0937+505 | 218 | 40.7 | 10.7 | 3.2E+13 | 8.5E+12 | 3.7E-08 | 9.8E-09 |
| WD1019-141 | 112 | 10.9 | | <8.7E+12 | | <2.0E-08 | |
| WD1024+326 | 400 | 51.8 | | <4.1E+13 | | <2.6E-08 | |
| WD1040+492 | 230 | 30.3 | 3.7 | 2.4E+13 | 2.9E+12 | 2.6E-08 | 3.2E-09 |
| WD1041+580 | 93 | 14.6 | | <1.2E+13 | | <3.1E-08 | |
| WD1057+719 | 141 | 26.3 | 2.9 | 2.1E+13 | 2.3E+12 | 3.7E-08 | 4.1E-09 |
| WD1109-225 | 82 | 13.8 | 4.0 | 1.1E+13 | 3.2E+12 | 3.4E-08 | 9.7E-09 |
| WD1234+481 | 129 | 10.4 | 3.0 | 8.3E+12 | 2.4E+12 | 1.6E-08 | 4.6E-09 |
| WD1254+223 | 67 | 15.1 | 5.0 | 1.2E+13 | 4.0E+12 | 4.5E-08 | 1.5E-08 |
| WD1302+597 | 79 | 17.4 | | <1.4E+13 | | <4.4E-08 | |
| WD1314+293 | 68 | 10.5 | 1.8 | 8.4E+12 | 1.4E+12 | 3.1E-08 | 5.3E-09 |
| WD1337+701 | 104 | 113.4 | | <9.1E+13 | | <2.2E-07 | |
| WD1440+753 | 98 | 11.9 | | <9.5E+12 | | <2.4E-08 | |
| WD1528+487 | 140 | 23.8 | 4.1 | 1.9E+13 | 3.3E+12 | 3.4E-08 | 5.8E-09 |
| WD1550+130 | 756 | 23.0 | | <1.8E+13 | | <6.1E-09 | |
| WD1603+432 | 114 | 16.1 | | <1.3E+13 | | <2.8E-08 | |
| WD1615-154 | 55 | 12.0 | | <9.6E+12 | | <4.4E-08 | |
| WD1620+647 | 174 | 17.2 | | <1.4E+13 | | <2.0E-08 | |
| WD1631+781 | 67 | 6.9 | 2.1 | 5.5E+12 | 1.6E+12 | 2.0E-08 | 6.2E-09 |
| WD1635+529 | 123 | 16.3 | | <1.3E+13 | | <2.7E-08 | |
| WD1636+351 | 109 | 15.1 | | <1.2E+13 | | <2.8E-08 | |
| WD1648+407 | 200 | 21.9 | 9.3 | 1.7E+13 | 7.4E+12 | 2.2E-08 | 9.3E-09 |
| WD1725+586 | 123 | 53.1 | 5.2 | 4.2E+13 | 4.1E+12 | 8.6E-08 | 8.4E-09 |
| WD1800+685 | 159 | 11.7 | 2.1 | 9.4E+12 | 1.7E+12 | 1.5E-08 | 2.7E-09 |
| WD1819+580 | 103 | 13.7 | | <1.1E+13 | | <2.7E-08 | |
| WD1844-223 | 62 | 7.4 | | <5.9E+12 | | <2.4E-08 | |
| WD1845+683 | 125 | 32.2 | | <2.6E+13 | | <5.2E-08 | |
| WD1917+509 | 105 | 12.5 | | <9.9E+12 | | <2.4E-08 | |
| WD1921-566 | 110 | 22.2 | | <1.8E+13 | | <4.0E-08 | |
| WD1950-432 | 140 | 17.5 | 5.0 | 1.4E+13 | 3.9E+12 | 2.5E-08 | 7.1E-09 |
| WD2004-605 | 58 | 9.4 | 2.8 | 7.5E+12 | 2.2E+12 | 3.3E-08 | 9.6E-09 |
| WD2014-575 | 51 | 29.0 | | <2.3E+13 | | <1.1E-07 | |

TABLE 5 - continued

| WD No. | d(pc) | ew | σ | $N_{OVI}$ (cm$^{-2}$) | Error | $n_{OVI}$ (cm$^{-3}$) | Error |
|---|---|---|---|---|---|---|---|
| WD2020-425 | 52 | 42.0 | | <3.4E+13 | | <1.6E-07 | |
| WD2116+736 | 177 | 9.9 | 2.6 | 7.9E+12 | 2.1E+12 | 1.1E-08 | 2.9E-09 |
| WD2124+191 | 46 | 44.7 | | <3.6E+13 | | <1.9E-07 | |
| WD2124-224 | 224 | 20.1 | 4.7 | 1.6E+13 | 3.7E+12 | 1.8E-08 | 4.2E-09 |
| WD2152-548 | 129 | 90.0 | 6.6 | 7.2E+13 | 5.2E+12 | 1.4E-07 | 1.0E-08 |
| WD2257-073 | 89 | 19.1 | 6.2 | 1.5E+13 | 4.9E+12 | 4.3E-08 | 1.4E-08 |
| WD2309+105 | 79 | 5.1 | | <4.1E+12 | | <1.3E-08 | |
| WD2321-549 | 192 | 22.3 | 6.3 | ~1.8E+13 | | ~2.3E-08 | |
| WD2331-475 | 82 | 6.6 | 2.4 | ~5.3E+12 | | ~1.6E-08 | |
| WD2350-706 | 92 | 22.6 | 3.6 | ~1.8E+13 | | ~4.9E-08 | |

WE ALSO GIVE THE ESTIMATED DISTANCE TO EACH STAR FROM WHICH THE VOLUME DENSITIES ARE CALCULATED. INCLUDED IN THE SAMPLE ARE THE UPPER LIMITS (3 SIGMA) ON COLUMN DENSITY FOR THOSE WHITE DWARFS WHERE INTERSTELLAR MATERIAL IS NOT DETECTED BUT WHERE THE SPECTRA ARE OF SUFFICIENT SIGNAL-TO-NOISE TO OBTAIN MEANINGFUL LIMITS ON THE AMOUNT OF O VI PRESENT. BLANK SPACES IN THE ERROR COLUMNS DENOTE THAT A DENSITY VALUE IS AN UPPER LIMIT. ~ INDICATES THAT THERE IS SIGNIFICANT SYSTEMATIC UNCERTAINTY IN THIS MEASUREMENT DUE TO THE BLENDING OF INTERSTELLAR AND PHOTOSPHERIC COMPONENTS.

TABLE 6
Key for figure 8

| WD | Star | Vis inspection | Distance | L | b | $N_{OVI}$ | $\log(N_{OVI})$ | Map index |
|---|---|---|---|---|---|---|---|---|
| WD1844-223 | REJ, EUVEJ | ND | 62.00 | 12.5022 | -9.2474 | <5.9e12 | <12.77 | 1 |
| WD2124-224 | REJ, EUVEJ | I | 224.00 | 27.3595 | -43.7596 | 1.6e13 | 13.20 | 2 |
| WD1314+293 | HZ43 | IR | 68.00 | 54.1058 | 84.1621 | 8.4e12 | 12.92 | 3 |
| WD1648+407 | REJ, EUVEJ | I | 200.00 | 64.6411 | 39.6014 | 1.7e13 | 13.23 | 4 |
| WD2257-073 | HD217411 | IR | 88.50 | 65.1744 | -56.9335 | 1.5e13 | 13.18 | 5 |
| WD1528+487 | REJ, EUVEJ | I | 140.00 | 78.8701 | 52.7205 | 1.9e13 | 13.28 | 6 |
| WD1725+586 | LB335 | IR | 123.20 | 87.1717 | 33.8299 | 4.2e13 | 13.62 | 7 |
| WD2309+105 | GD246 | ND | 79.00 | 87.2625 | -45.1174 | <4.1e12 | <12.61 | 8 |
| WD1800+685 | KUV | I | 159.00 | 98.73 | 29.7751 | 9.4e12 | 12.97 | 9 |
| WD2116+736 | KUV | I | 177.00 | 109.3900 | 16.9240 | 7.9e12 | 12.90 | 10 |
| WD1631+781 | REJ, EUVEJ | I | 67.00 | 111.2953 | 33.5782 | 5.5e12 | 12.74 | 11 |
| WD0004+330 | GD2 | I | 98.80 | 112.4803 | -28.6882 | 4.9e12 | 12.69 | 12 |
| WD1234+481 | PG | I | 129.00 | 129.8100 | 69.0086 | 8.3e12 | 12.92 | 13 |
| WD1057+719 | PG | IR | 141.00 | 134.4790 | 42.9215 | 2.1e13 | 13.32 | 14 |
| WD1040+492 | REJ, EUVEJ | IR | 230.00 | 162.6693 | 57.0087 | 2.4e13 | 13.38 | 15 |
| WD0937+505 | PG | I | 218.00 | 166.9021 | 47.1203 | 3.2e13 | 13.51 | 16 |
| WD0235-125 | PHL1400 | ND | 86.20 | 187.3990 | -61.1220 | <1.6e13 | <13.20 | 17 |
| WD0346-011 | GD50 | ND | 29.00 | 188.9534 | -40.0974 | <4.5e12 | <12.65 | 18 |
| WD0549+158 | GD71 | ND | 49.00 | 192.0286 | -5.3382 | <3.2e12 | <12.51 | 19 |
| WD0455-282 | REJ, EUVEJ | IB+P | 102.00 | 229.2948 | -36.1650 | 9.0e12 | 12.95 | 20 |
| WD0229-481 | REJ, EUVEJ | ND | 239.60 | 266.6179 | -61.5914 | <9.6e12 | <12.98 | 21 |
| WD0320-539 | LB1663 | ND | 124.10 | 267.3001 | -51.6355 | <1.3e13 | <13.11 | 22 |
| WD0830-535 | REJ, EUVEJ | I | 82.00 | 270.1121 | -8.2658 | 2.5e13 | 13.40 | 23 |
| WD1109-225 | HD97277 | I | 81.56 | 274.7780 | 34.5360 | 1.1e13 | 13.04 | 24 |
| WD0715-704 | REJ, EUVEJ | I | 94.00 | 281.4000 | -23.5000 | 9.2e12 | 12.96 | 25 |
| WD0325-857 | REJ, EUVEJ | ND | 35.00 | 299.8575 | -30.6815 | <4.2e12 | <12.62 | 26 |
| WD0027-636 | REJ, EUVEJ | I+P | 236.00 | 306.9798 | -53.5497 | ~1.1e13 | ~13.04 | 27 |
| WD2350-706 | HD223816 | I+P | 92.00 | 309.9147 | -45.9375 | ~1.8e13 | ~13.26 | 28 |
| WD1254+223 | GD153 | IR | 67.00 | 317.2548 | 84.7465 | 1.2e13 | 13.08 | 29 |
| WD2321-549 | REJ, EUVEJ | I+P | 192.00 | 326.9080 | -58.2104 | ~1.8e13 | ~13.26 | 30 |
| WD2331-475 | REJ, EUVEJ | I+P | 82.00 | 334.8358 | -64.8080 | ~5.3e12 | ~12.72 | 31 |
| WD2004-605 | REJ, EUVEJ | I | 58.00 | 336.5813 | -32.8586 | 7.5e12 | 12.88 | 32 |
| WD2014-575 | LS210-114 | ND | 51.00 | 340.2003 | -34.2481 | <2.3e13 | <13.36 | 33 |
| WD1950-432 | HS | I | 140.00 | 356.5050 | -29.0920 | 1.4e13 | 13.15 | 34 |
| WD2020-425 | REJ, EUVEJ | ND | 52.20 | 358.3597 | -34.4540 | <3.4e13 | <13.53 | 35 |
| WD1615-154 | G153-41 | ND | 55.00 | 358.7916 | 24.1800 | <9.6e12 | <12.98 | 36 |

~ INDICATES THAT THERE IS SIGNIFICANT SYSTEMATIC UNCERTAINTY IN THIS MEASUREMENT DUE TO THE BLENDING OF INTERSTELLAR AND PHOTOSPHERIC COMPONENTS.

FIG 1. — *FUSE* spectrum of the hot DA white dwarf WD 0131-164 (GD 984), with the main photospheric and interstellar features labelled. The wavelengths of all these absorption lines can be obtained from Table 2. We note that the line identified as photospheric Si IV 1066.62Å in this plot may often be contaminated by interstellar Ar I, which occurs at the same wavelength.

FIG 2. — White Dwarf heliospheric velocity ($v_{phot}$) compared to the ISM velocity ($v_{ISM}$) for the measurements listed in Table 3.

FIG 3. — The number distribution of velocity differences ($v_{PHOT} - v_{ISM}$). The distribution has a mean of 20.3 km s$^{-1}$.

FIG 4. — Normalised profiles vs. the heliocentric velocity of the local ISM C II and O I absorption lines (lower panel) compared to the O VI region. The vertical dashed line marks the photospheric velocity and the dash-dot line the mean interstellar velocity for each star. We mark the statistical error bars in the top panel of each plot to illustrate the significance of the O VI detections we report but have left them out of the bottom panel for clarity.

FIG 5. — O VI column density as a function of stellar distance (pc). The filled red circles and error bars are measured values, while the filled blue circles and downward bar are 3σ upper limits. For the four stars that have blended ISM and photospheric O VI components, the error bars are not shown because of the larger and unquantifiable uncertainty in the measurement.

FIG 6. — O VI volume density as a function of stellar distance (pc). The filled red circles and error bars are measured values, while the filled blue circles and downward bar are 3σ upper limits. For the four stars that have blended ISM and photospheric O VI components, the error bars are not shown because of the larger and unquantifiable uncertainty in the measurement.

FIG 7. — Location of detections and non-detections of O VI absorption towards the hot white dwarfs in this study, 7 B-stars from Welsh & Lallement (2008) and 8 B-stars from Bowen et al. (2008) superimposed on a map of the spatial distribution of the SXRB emission intensity (Snowden et al 1997). Symbols are filled black circles = O VI detections below 80 pc; filled white circles = non-detections below 80 pc; filled black squares = detections between 80 and 120 pc; filled white squares = non-detections between 80 and 120pc. The colour bar to the right of the map is the key to the intensity of the soft X-ray background in units of 10$^{-6}$ counts s$^{-1}$ arcmin$^{-2}$.

FIG 8. — Density slices through the local ISM. White is low density and black is high density. Filled yellow circles represent detections of interstellar O VI while open circles mark non-detections. The stars numbers inside each symbol correspond to the list in Table 6. The log of the O VI column density is indicated in red beside each circle. The Galactic longitude of each slice is indicated in each panel along with the locations of the Galactic poles. The tick marks on each axis are intervals of 100pc distance from the Sun. We note that star 21 in the l=277° slice actually lies off the map in the direction of the arrow at a distance of 240pc.

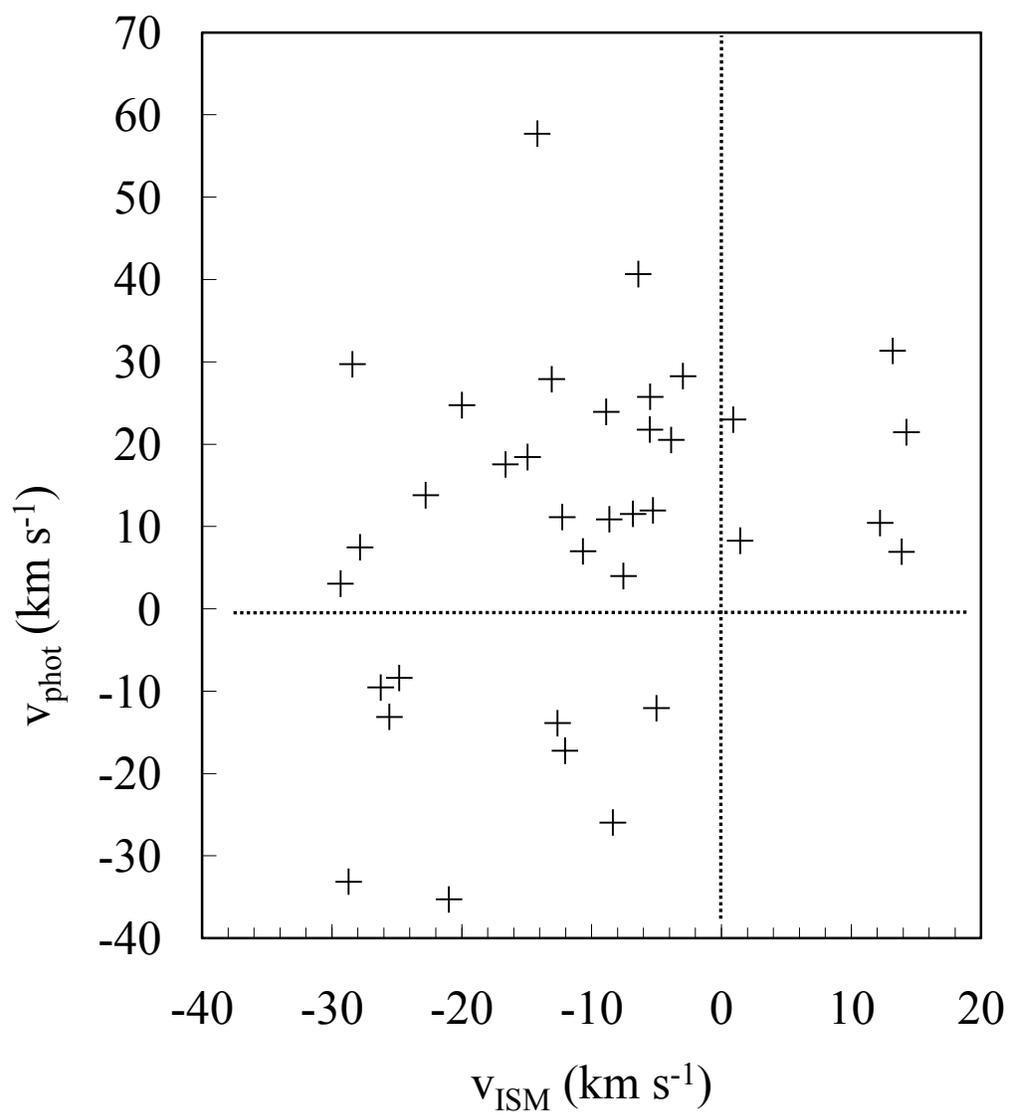

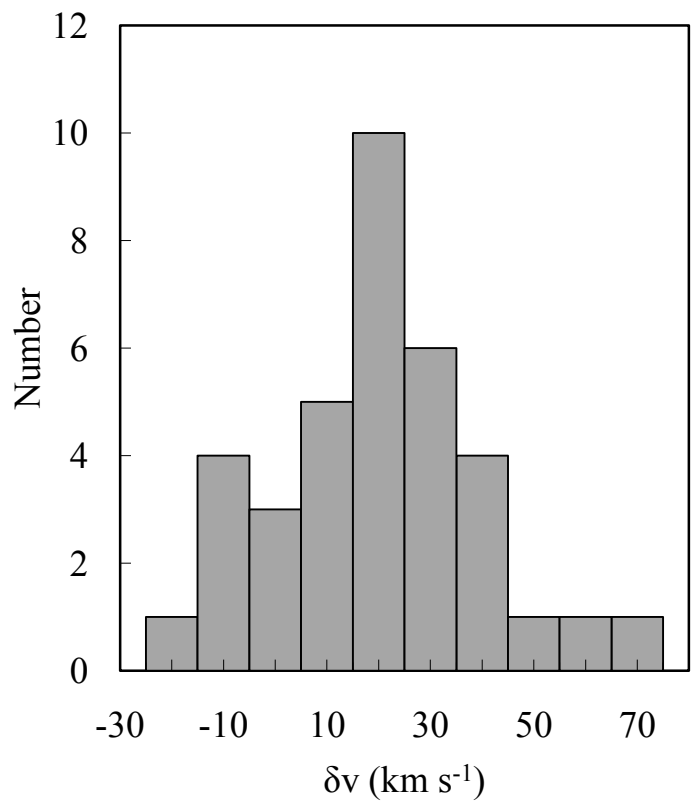

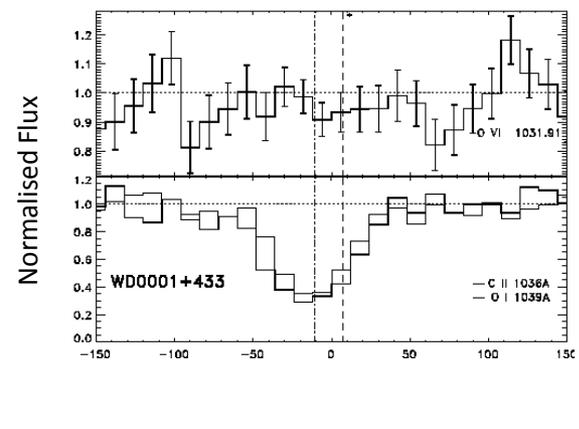 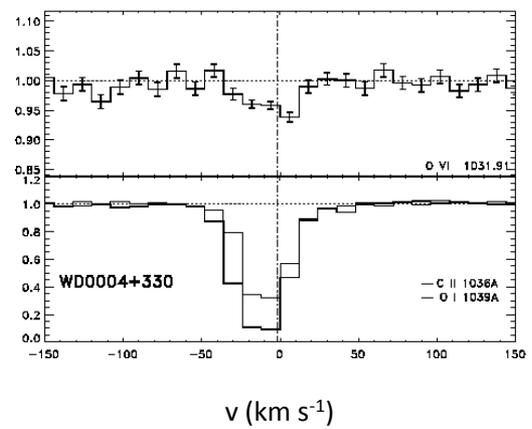 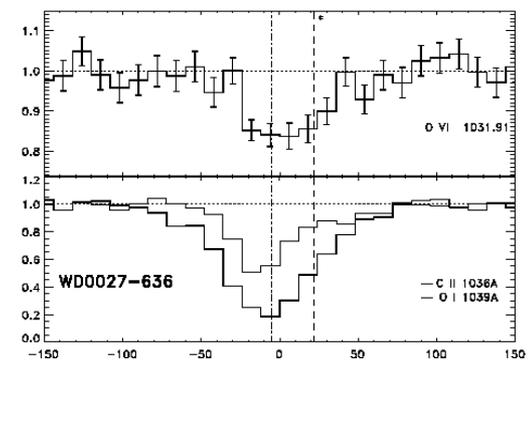
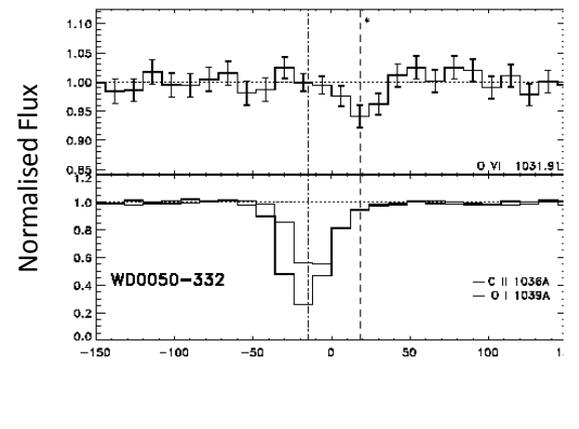 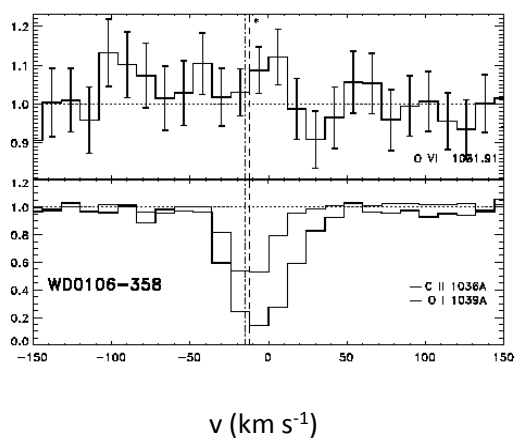 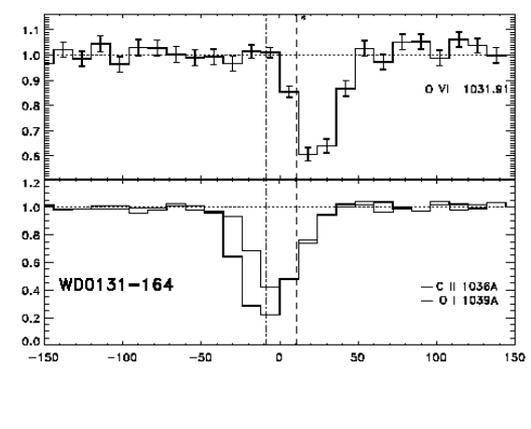
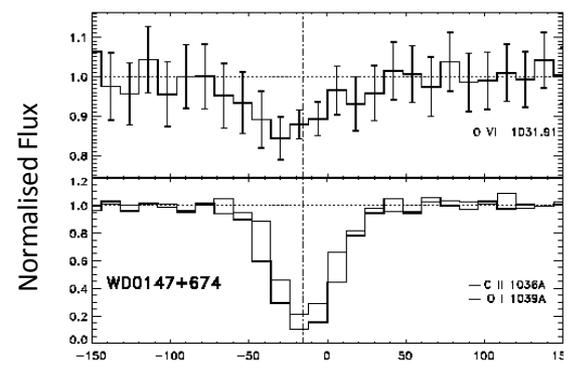 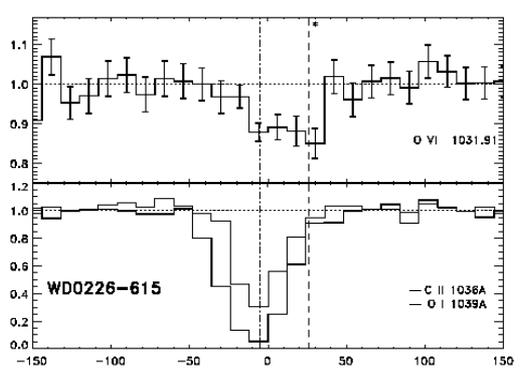 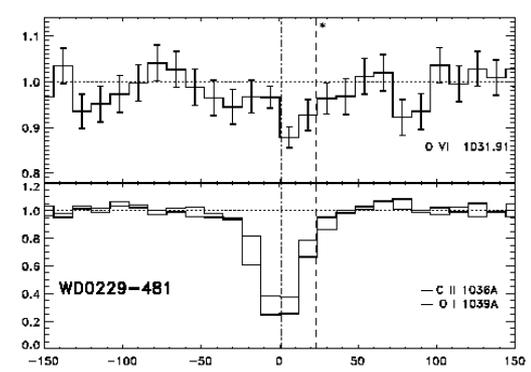

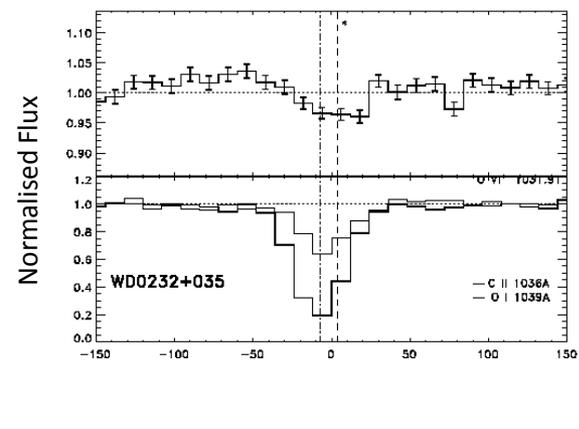
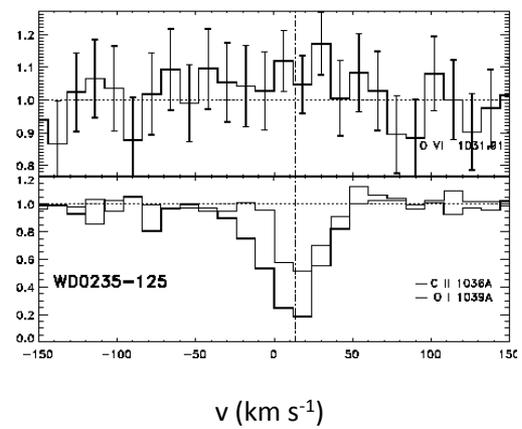
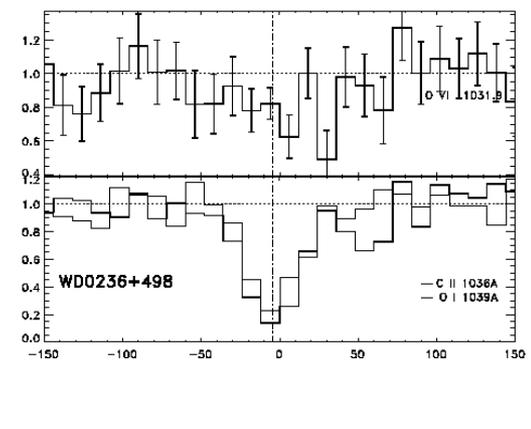
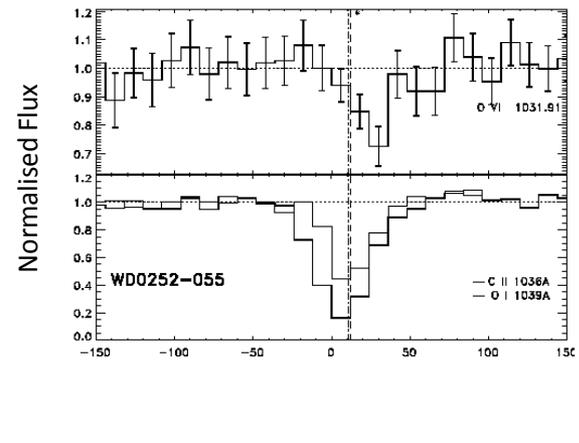
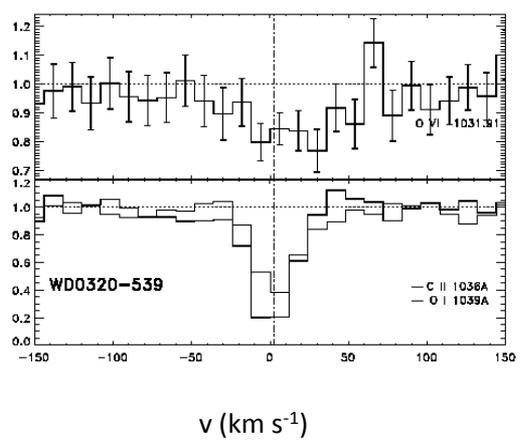
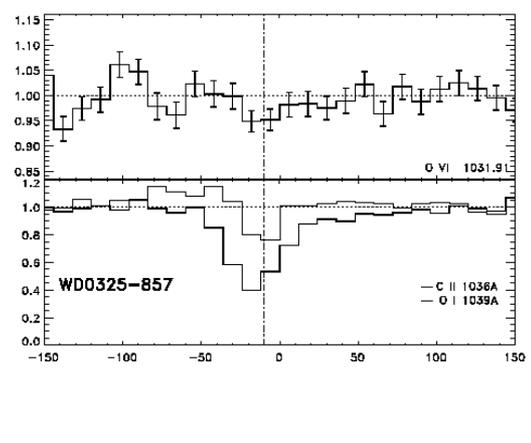
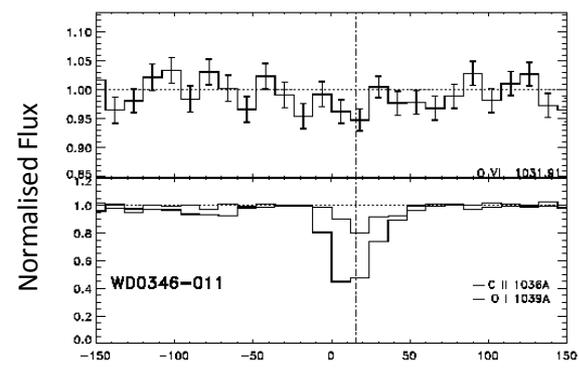
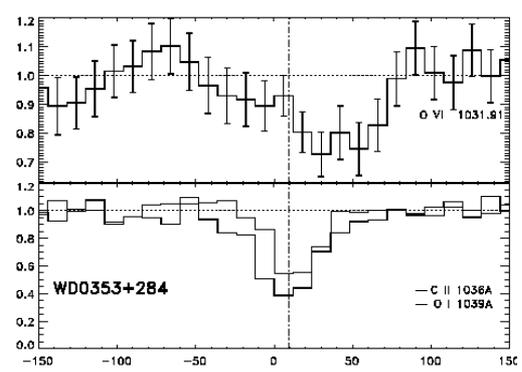
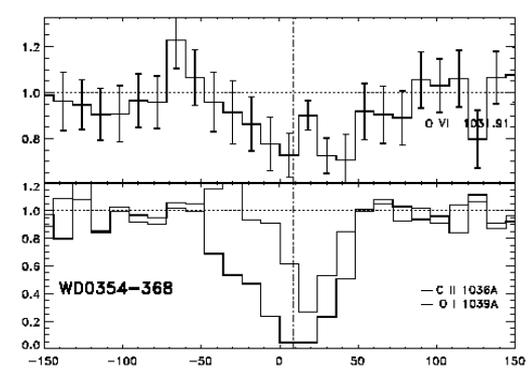

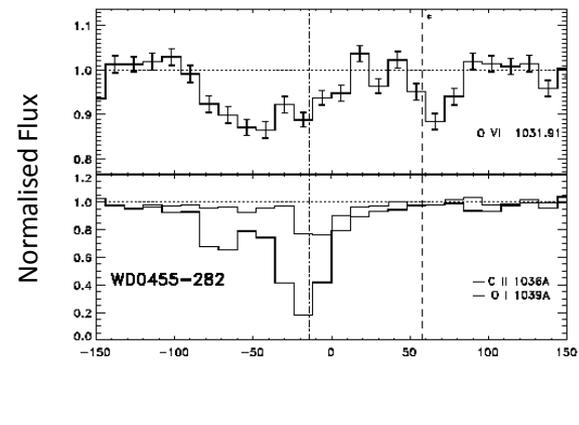
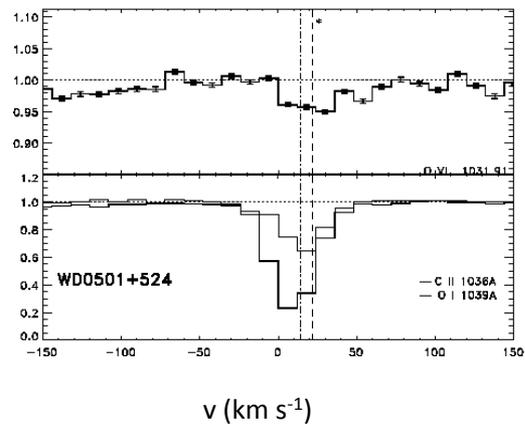
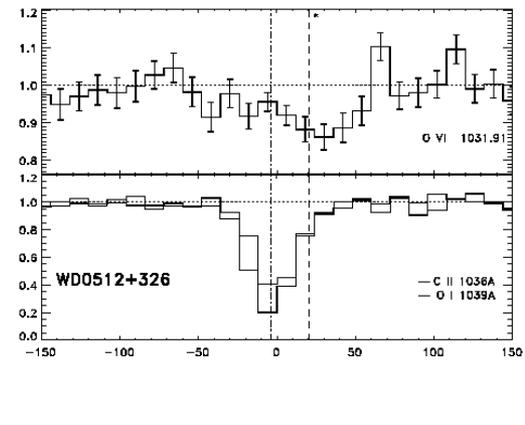
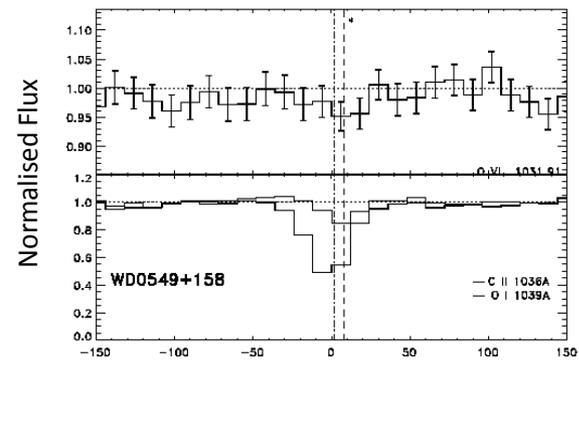
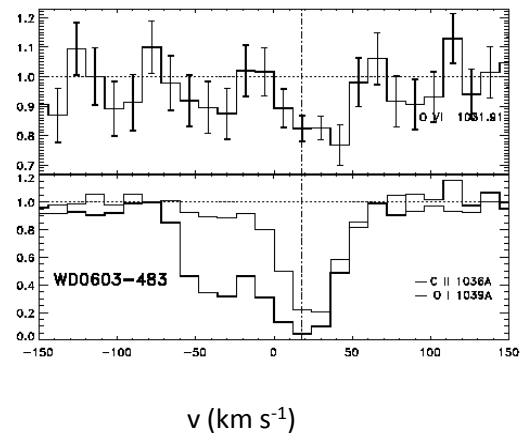
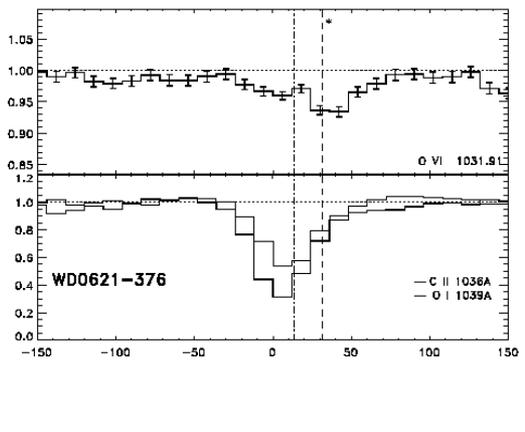
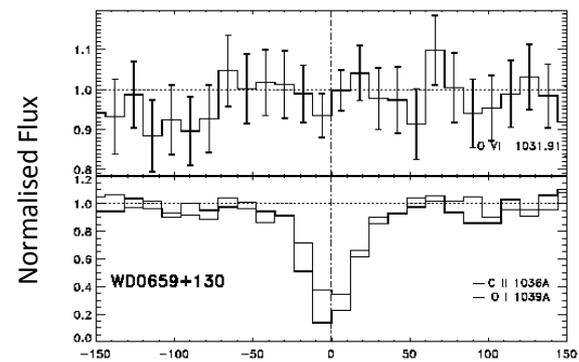
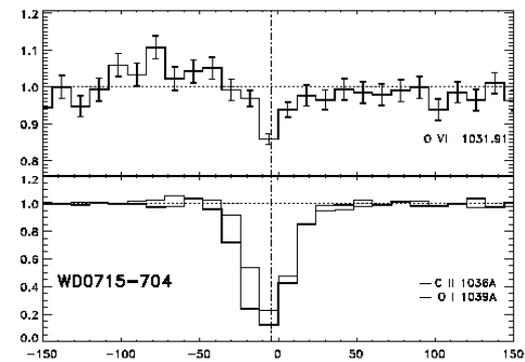
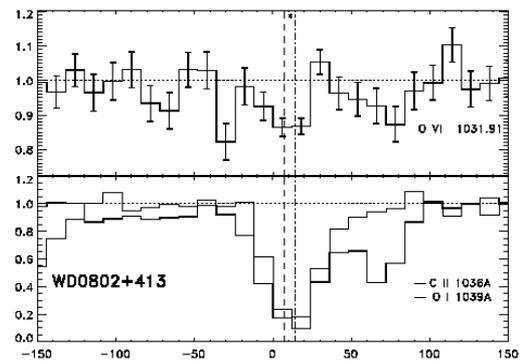

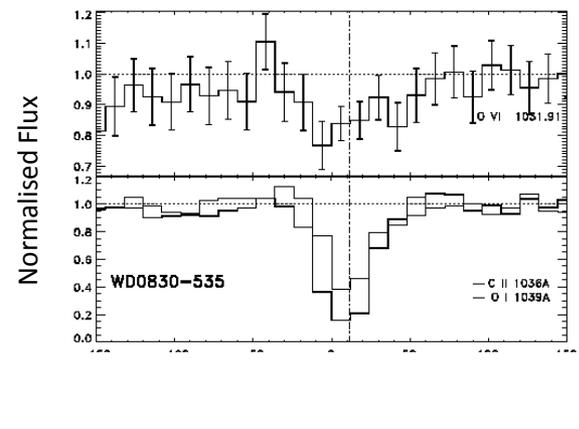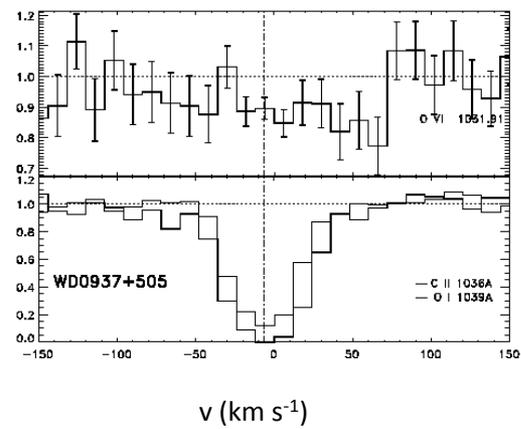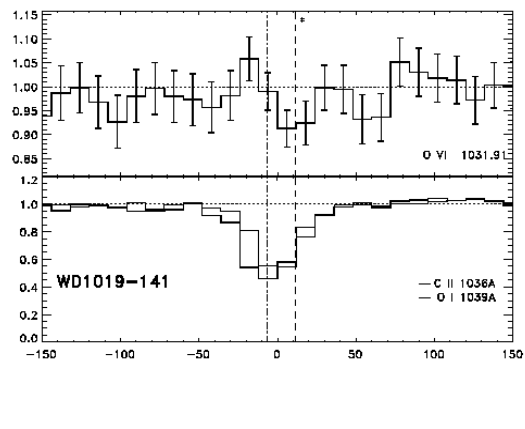
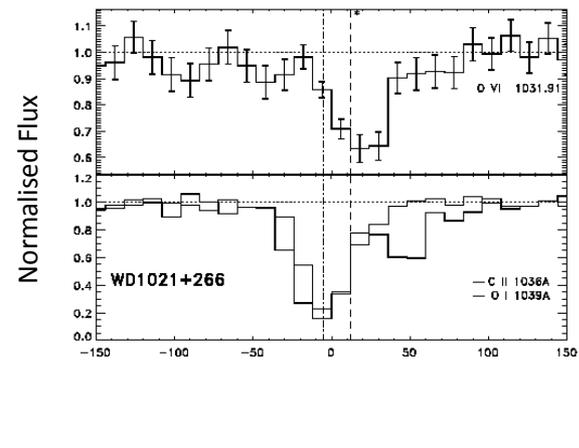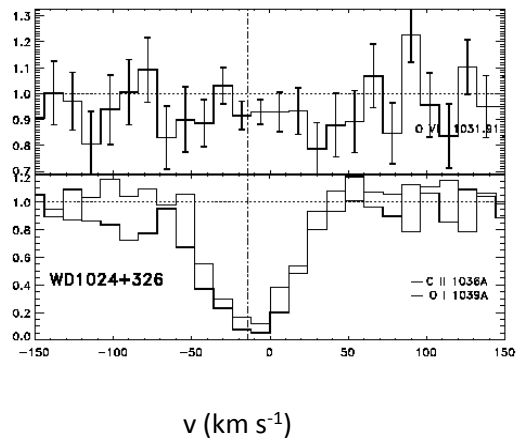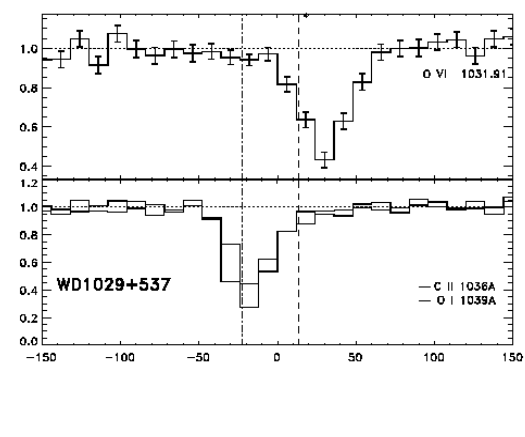
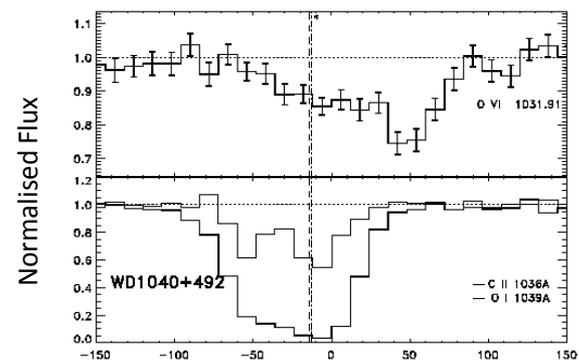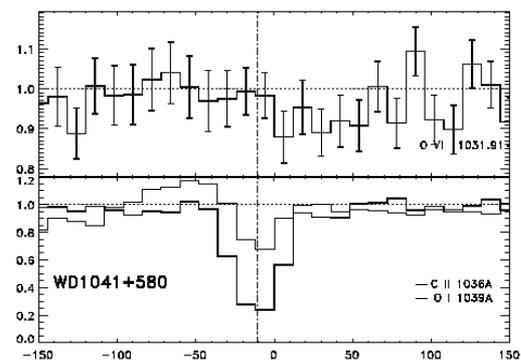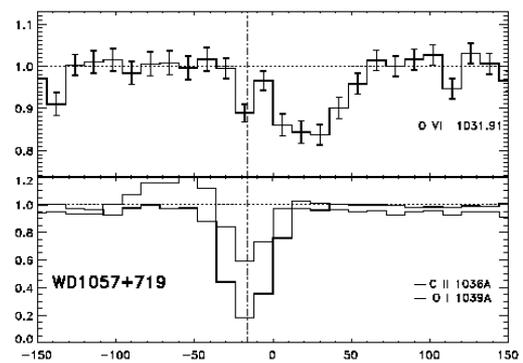

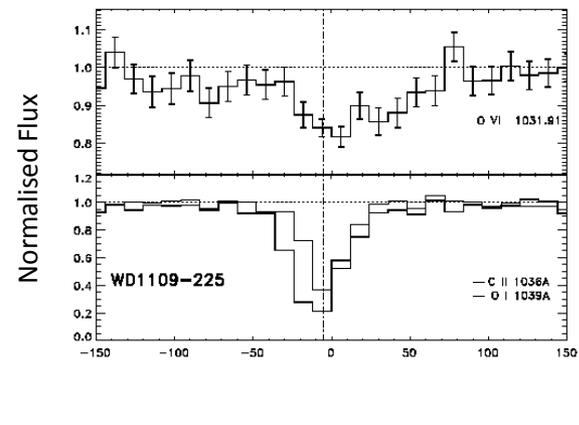
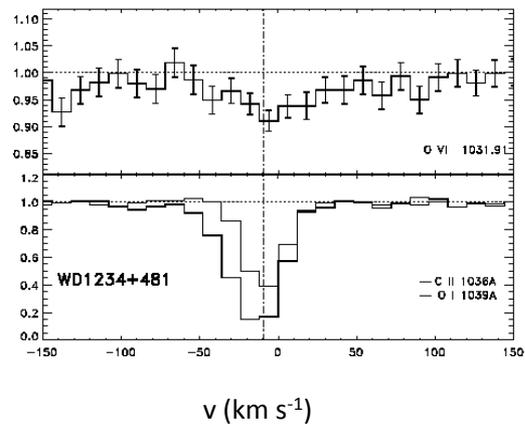
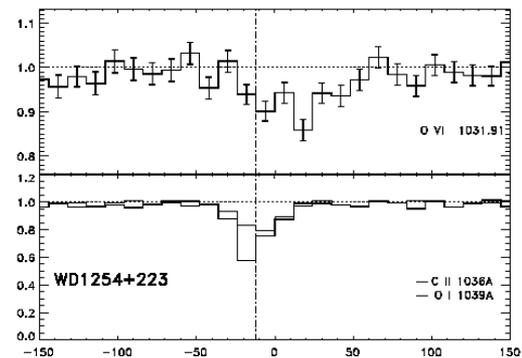
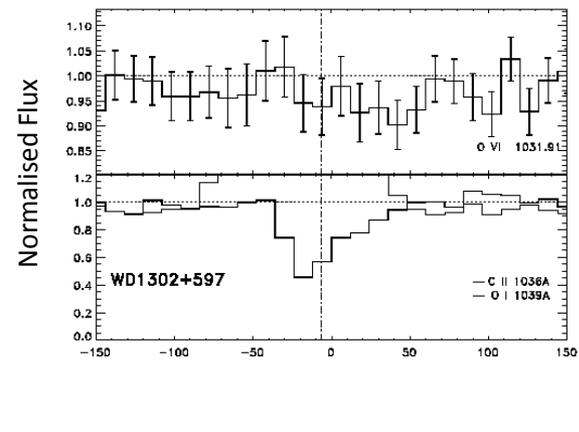
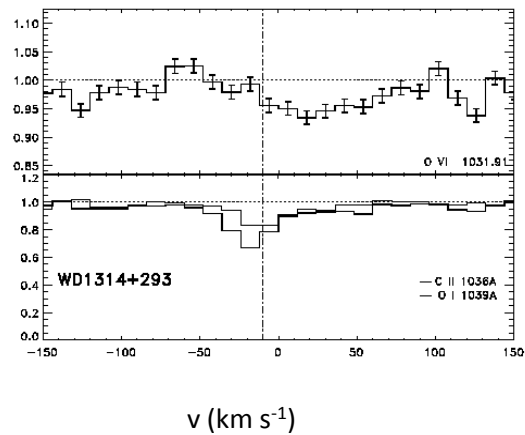
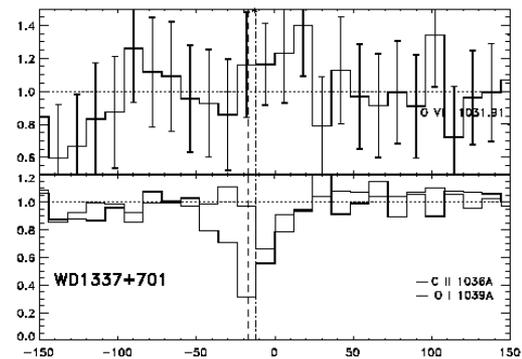
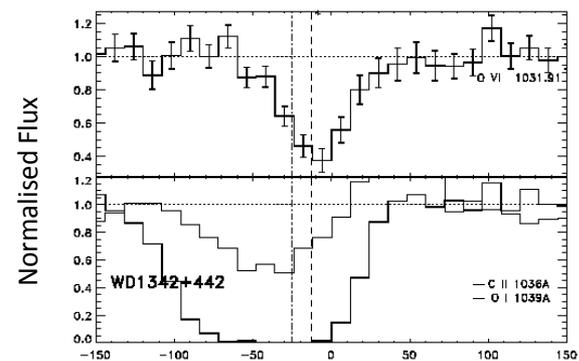
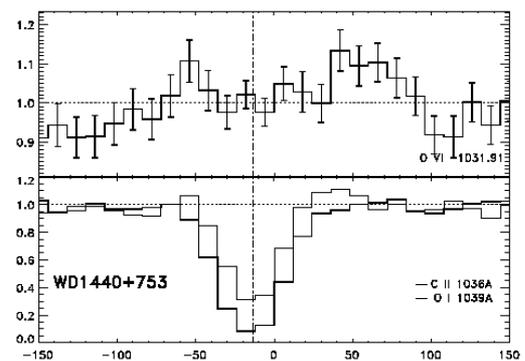
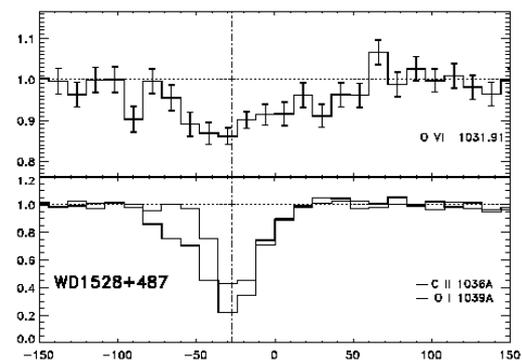

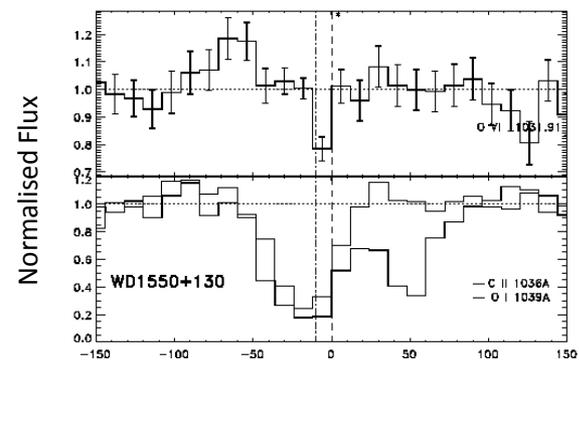
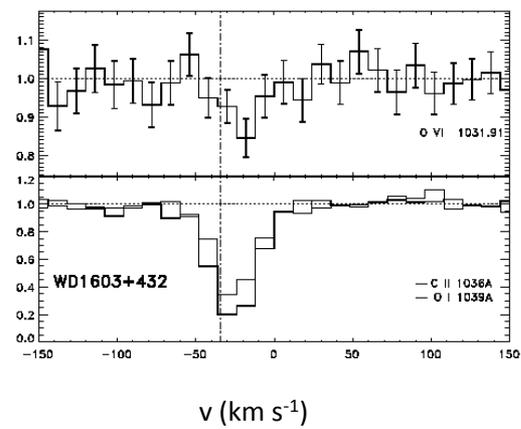
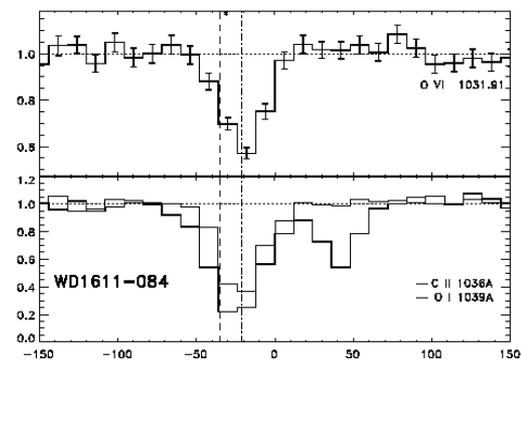
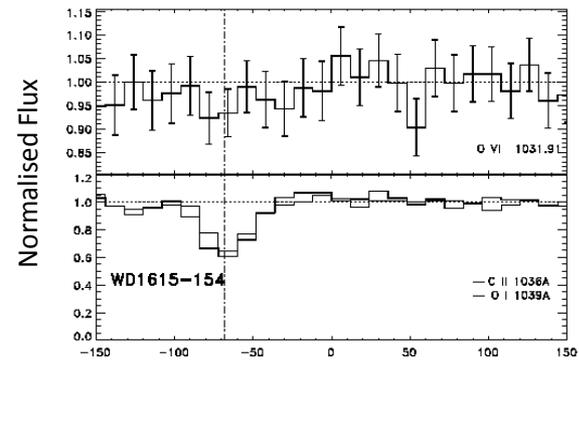
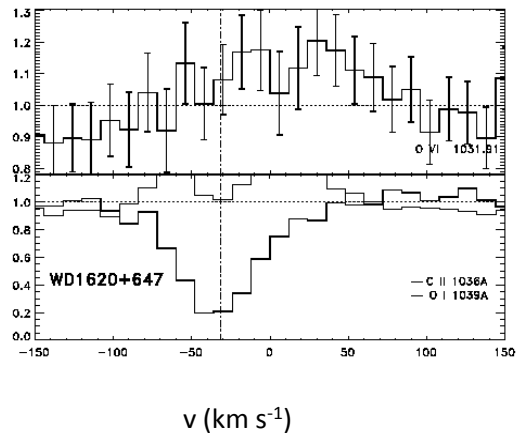
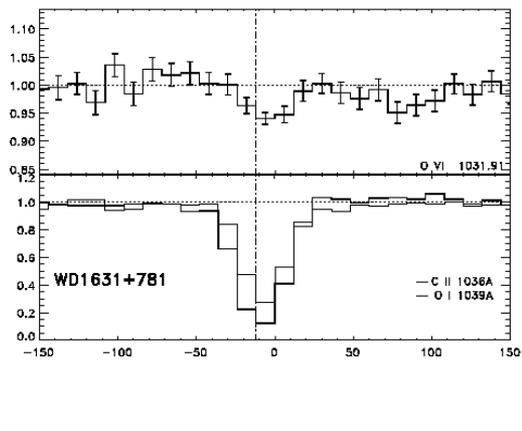
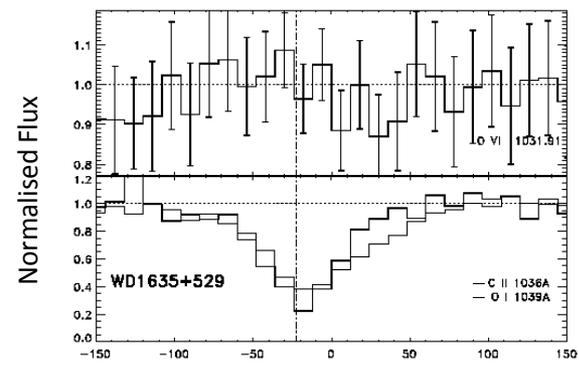
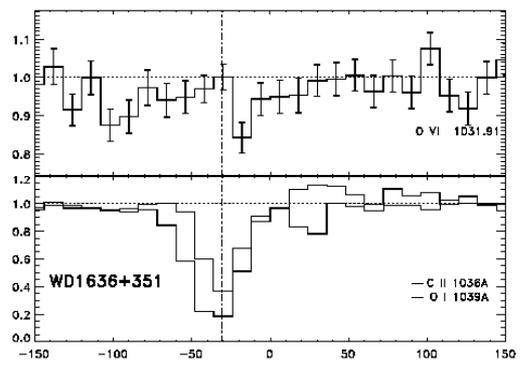
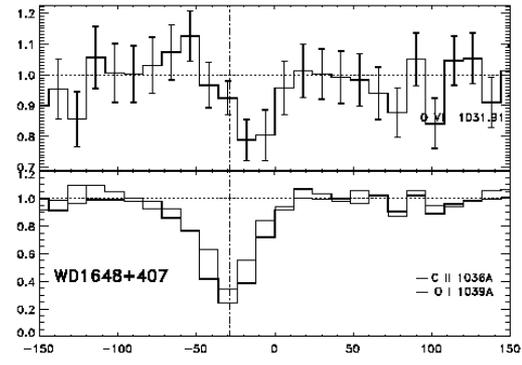

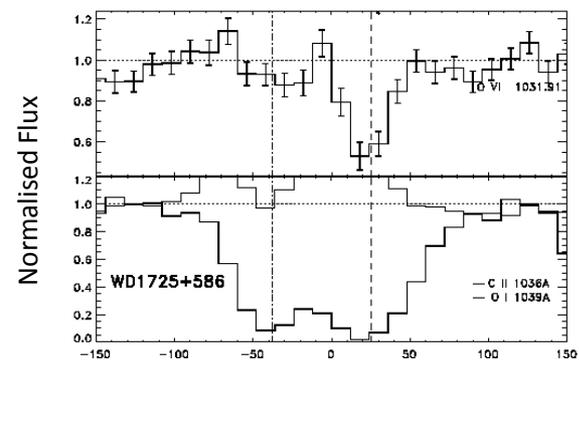
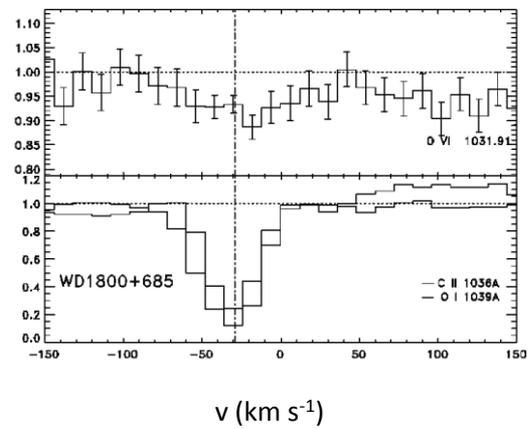
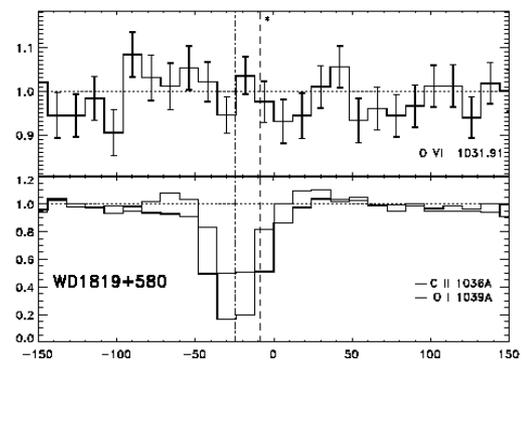
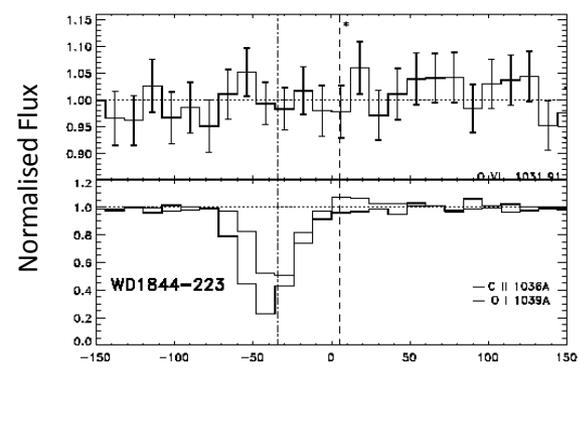
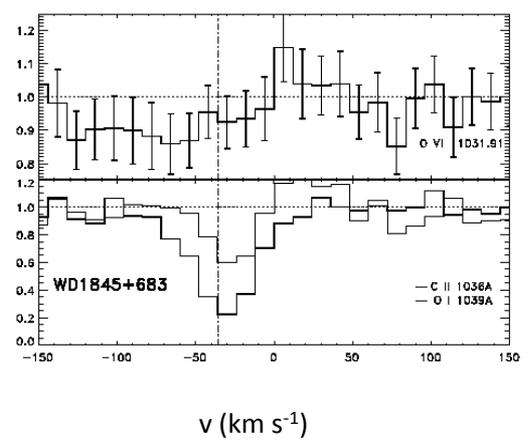
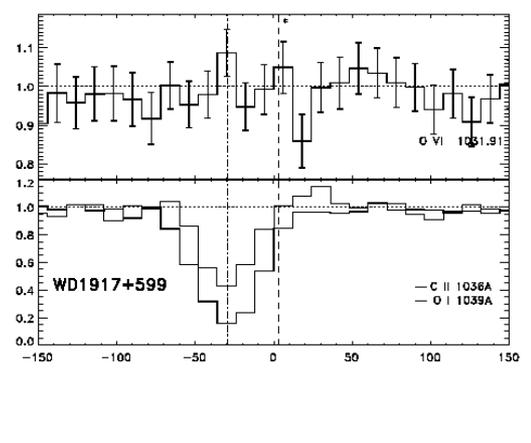
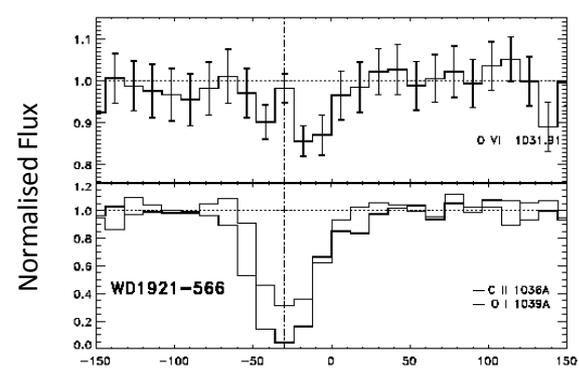
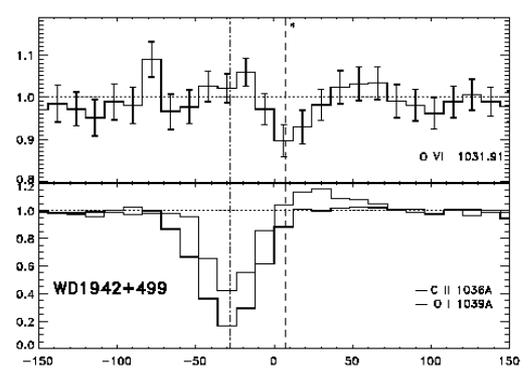
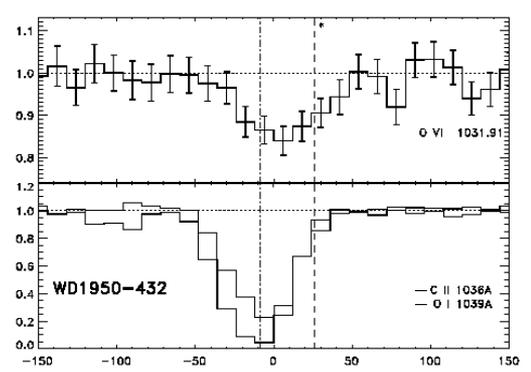

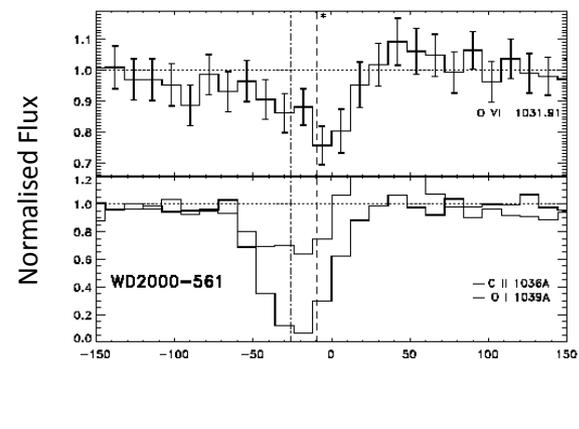
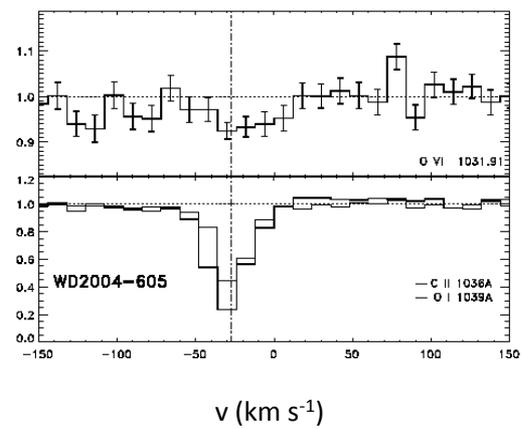
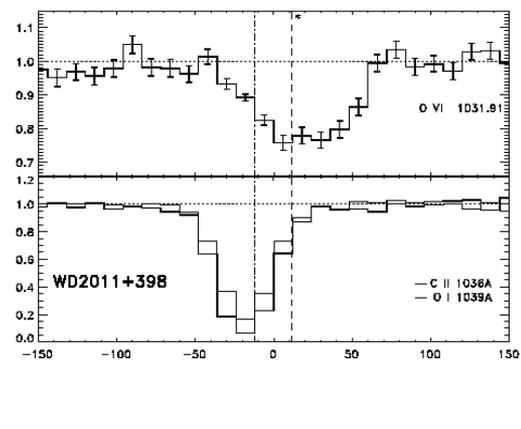
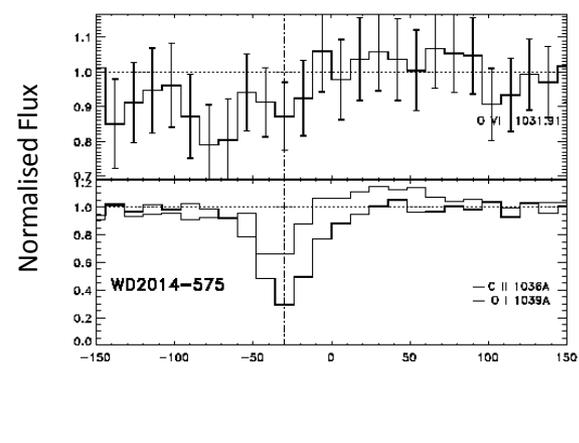
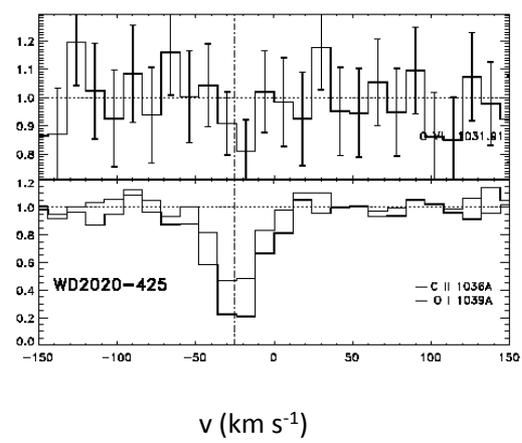
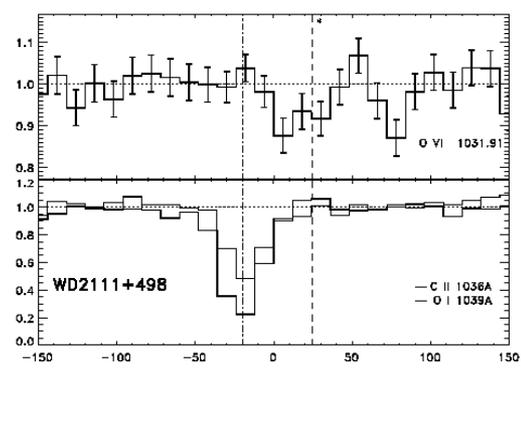
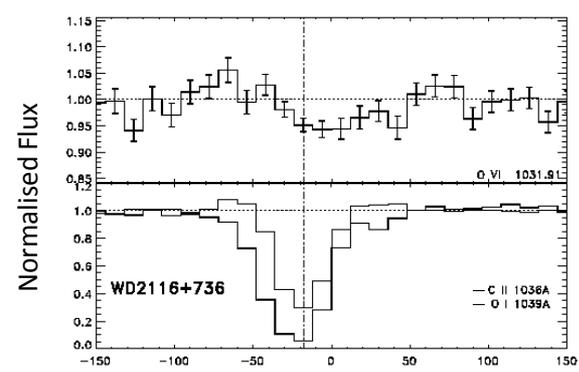
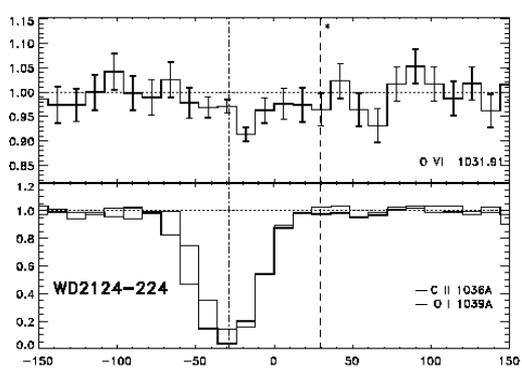
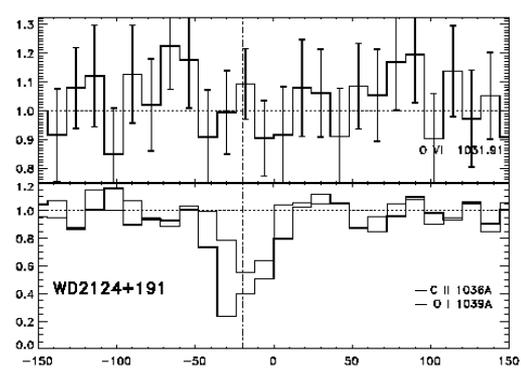

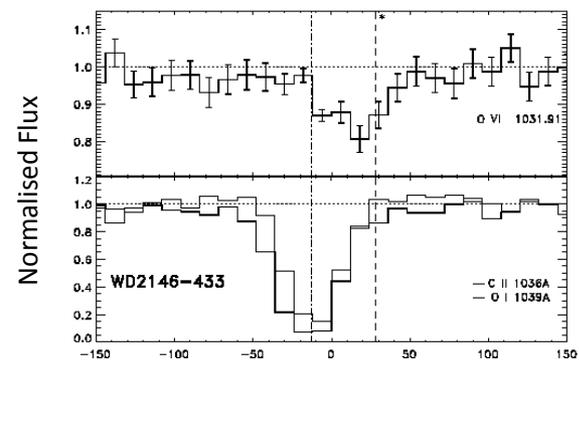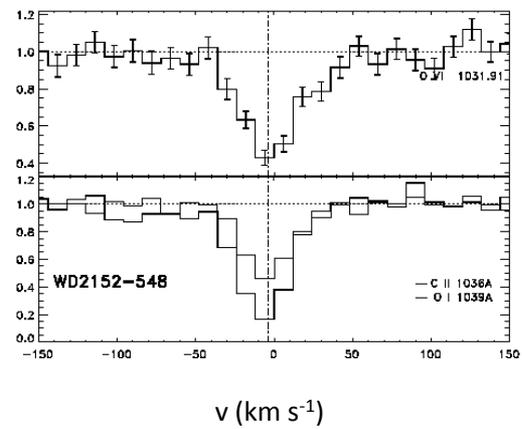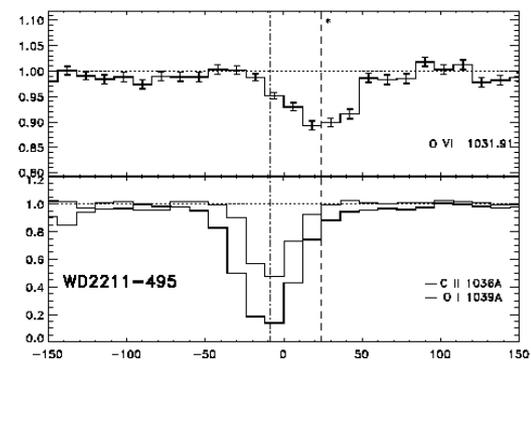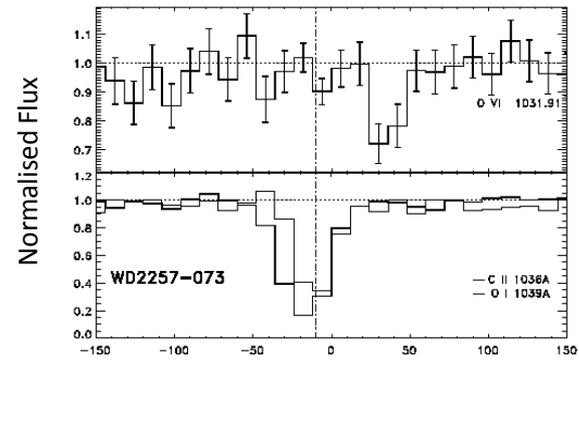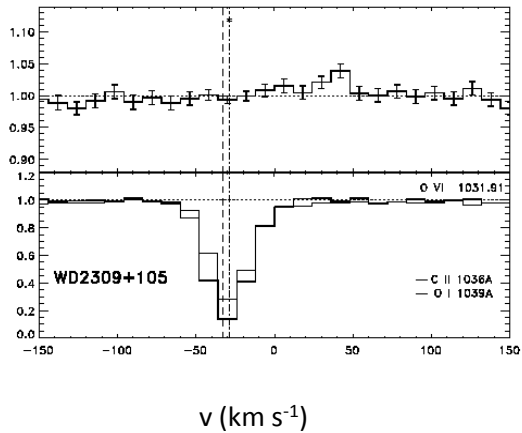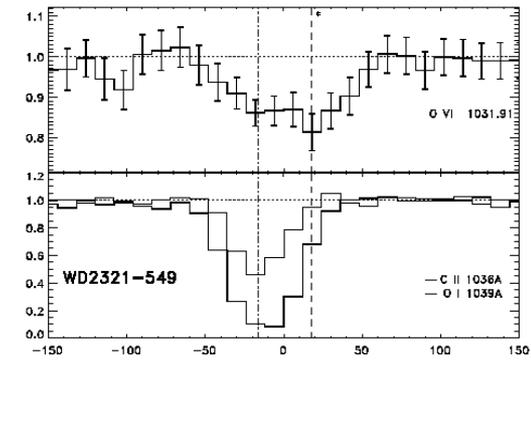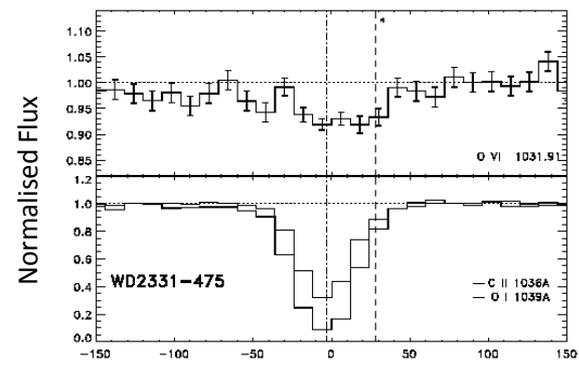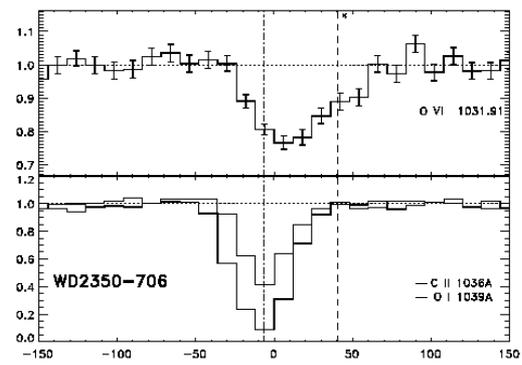

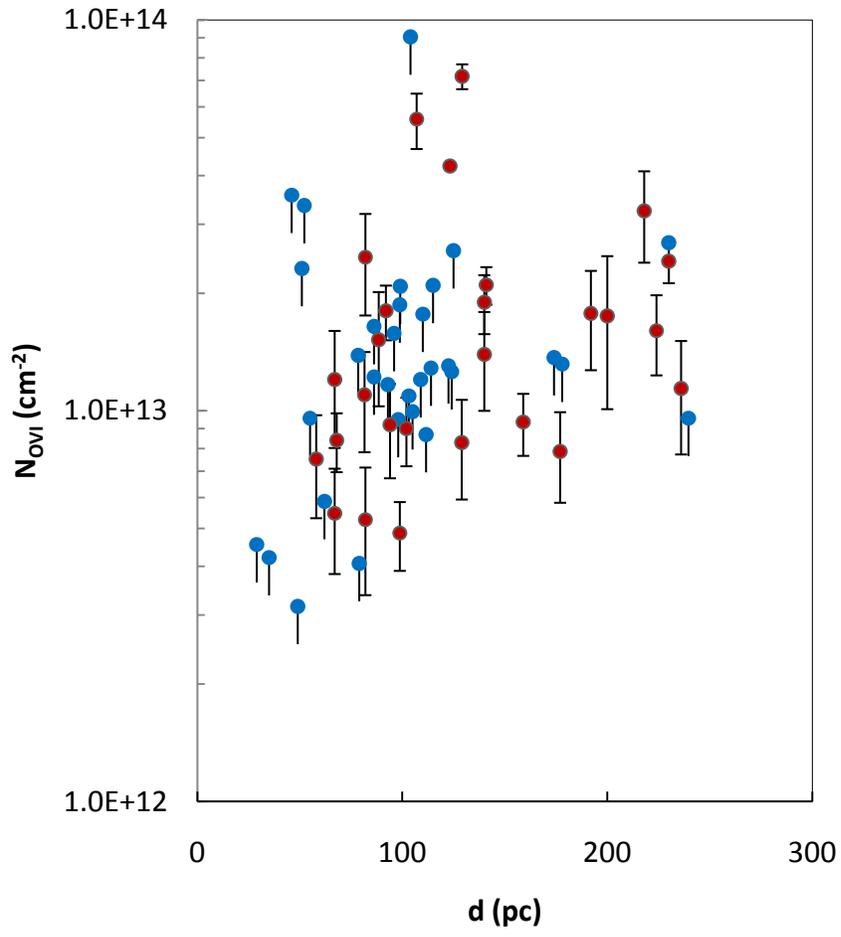

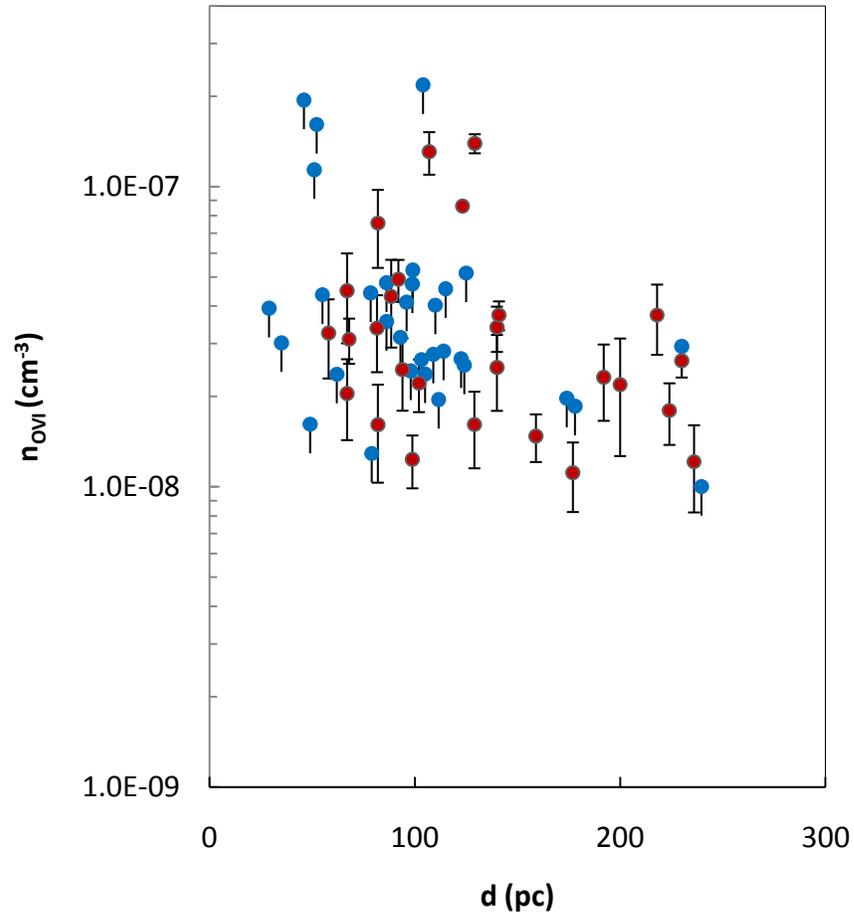

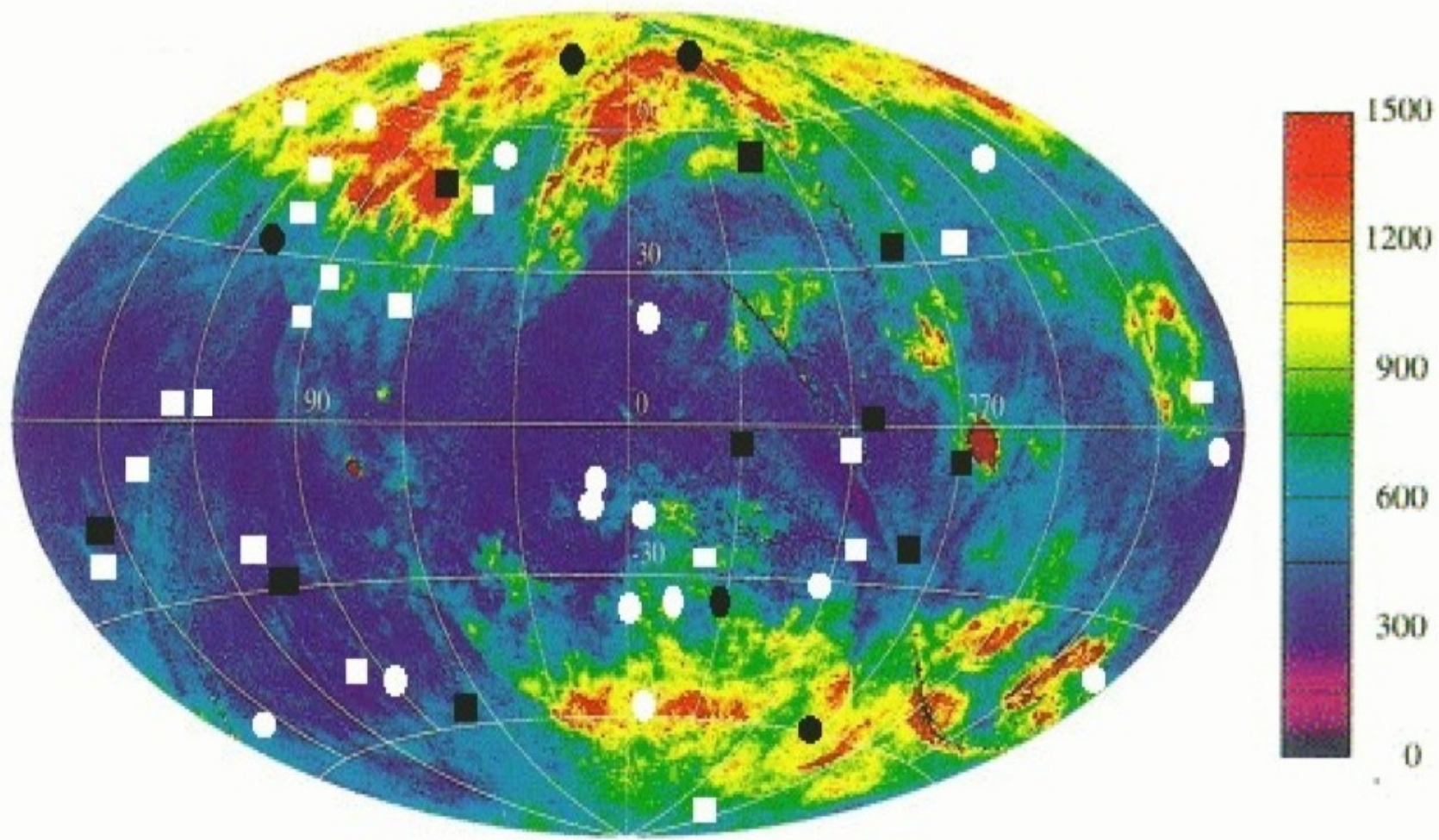

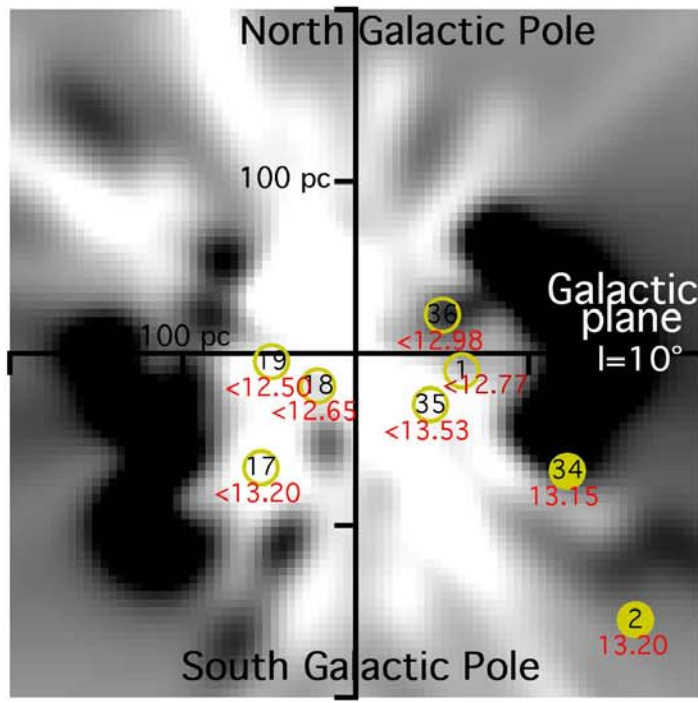
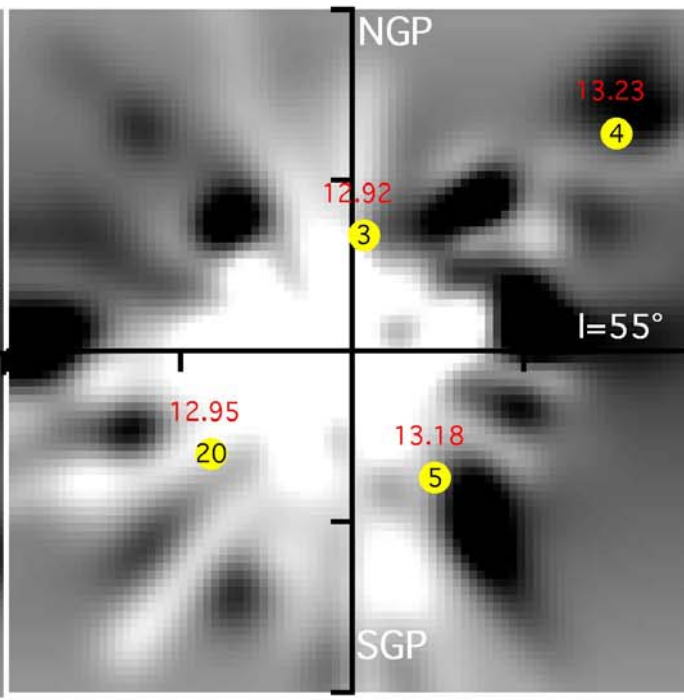
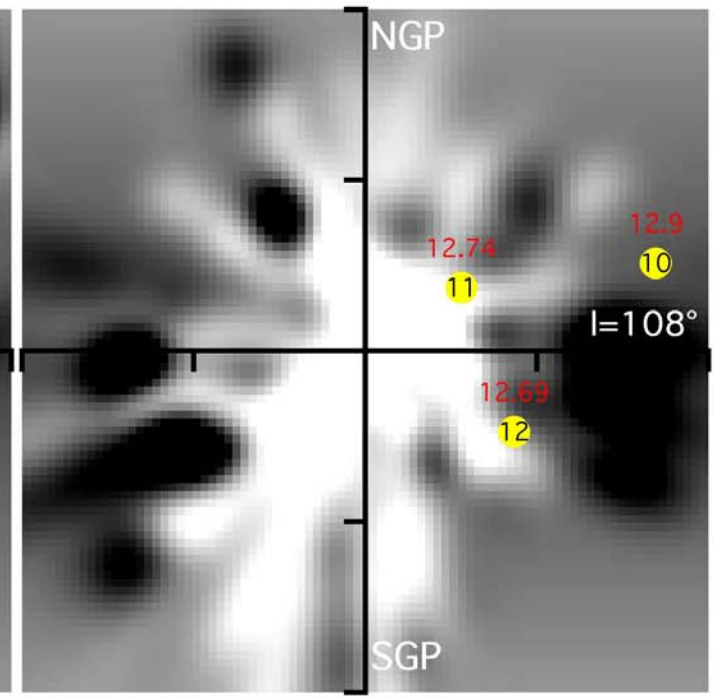
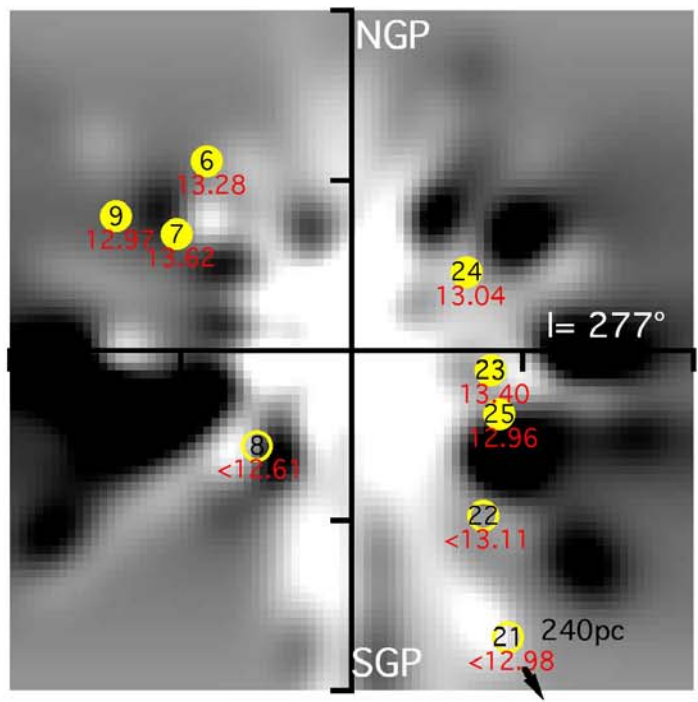
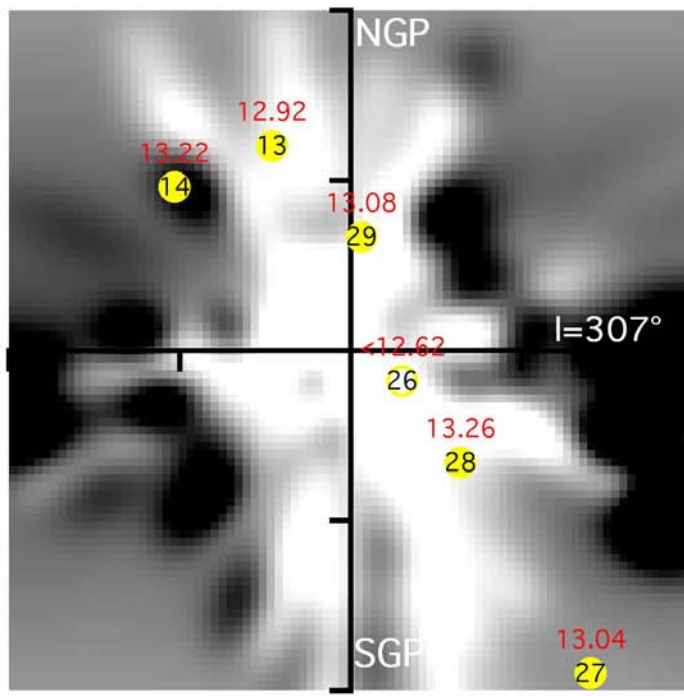
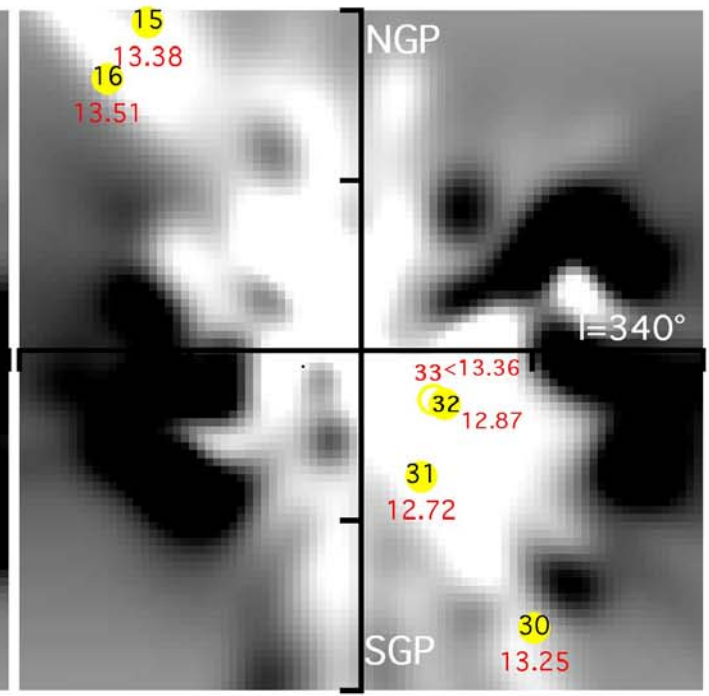